\documentclass[journal,twoside,web]{ieeecolor}
\usepackage{caption} 
\usepackage{xr} 
\usepackage{xcolor}
\usepackage{hyperref}
\usepackage{generic}
\usepackage{cite}
\usepackage{mathtools,amssymb,amsfonts}
\usepackage{subfig}
\usepackage{graphicx}
\usepackage{algorithm}
\usepackage{algpseudocode}
\usepackage{textcomp}
\usepackage{booktabs} 
\newcommand{\etal}{\textit{et al}. }

\usepackage{multirow}
\usepackage{comment}
\usepackage{threeparttable}
\usepackage{cite,etoolbox}
\usepackage{adjustbox}
\usepackage{array, makecell}

\def\BibTeX{{\rm B\kern-.05em{\sc i\kern-.025em b}\kern-.08em
    T\kern-.1667em\lower.7ex\hbox{E}\kern-.125emX}}
\begin{document}
\title{{M3D: Manifold-based Domain Adaptation with Dynamic Distribution for Non-Deep Transfer Learning in Cross-subject and Cross-session EEG-based Emotion Recognition}}
\author{Ting Luo\textsuperscript{1},
Jing Zhang\textsuperscript{1},
Yingwei Qiu,
Li Zhang,Yaohua Hu,
Zhuliang Yu, and Zhen Liang\textsuperscript{*}
\thanks{This work was supported by the National Natural Science Foundation of China under Grant 62276169, the Medical-Engineering Interdisciplinary Research Foundation of Shenzhen University, the Shenzhen-Hong Kong Institute of Brain Science-Shenzhen Fundamental Research Institutions under Grant 2022SHIBS0003, the STI 2030-Major Projects 2021ZD0200500, and the Shenzhen Science and Technology Program (No. JCYJ20241202124222027
and JCYJ20241202124209011).}
\thanks{Ting Luo, Li Zhang, and Zhen Liang are with the School of Biomedical Engineering, Medical School, Shenzhen University, Shenzhen 518060, China, also with the Guangdong Provincial Key Laboratory of Biomedical Measurements and Ultrasound Imaging, Shenzhen 518060, China. E-mail: 2310247018@email.szu.edu.cn, and \{lzhang, janezliang\}@szu.edu.cn.}
\thanks{Jing Zhang is with the School of Public Health, Shenzhen University Medical School, Shenzhen, China. E-mail: {zhangjing1985zj@163.com.}}
\thanks{Yingwei Qiu is Department of Radiology, Nanshan Hospital of Shenzhen University, Shenzhen, China. E-mail: qiuyw1201@gmail.com.}
\thanks{Yaohua Hu is with the School of Mathematical Sciences, Shenzhen University, Shenzhen, China. E-mail: mayhhu@szu.edu.cn.}
\thanks{Zhuliang Yu is with Shien-Ming Wu School of Intelligent Engineering, South China University of Technology, Guangzhou, Guangdong, China, and also with Institute for Super Robotics, Guangzhou, Guangdong, China. E-mail: zlyu@scut.edu.cn.}
\thanks{\textsuperscript{*}Corresponding authors: Zhen Liang.}}

\maketitle
\footnotetext[1]{\hspace{1mm}Equal contributions.}

\begin{abstract}
Emotion decoding using Electroencephalography (EEG)-based affective brain-computer interfaces (aBCIs) is crucial for affective computing but is hindered by EEG’s non-stationarity, individual variability, and the high cost of large-scale labeled data. Deep learning-based approaches, while effective, require substantial computational resources and large datasets, limiting their practicality. To address these challenges, we propose Manifold-based Domain Adaptation with Dynamic Distribution (M3D), a lightweight non-deep transfer learning framework. M3D includes four main modules: manifold feature transformation, dynamic distribution alignment, classifier learning, and ensemble learning. The data undergoes a transformation onto an optimal Grassmann manifold space, enabling dynamic alignment of the source and target domains. This process prioritizes both marginal and conditional distributions according to their significance, ensuring enhanced adaptation efficiency across various types of data. In the classifier learning, the principle of structural risk minimization is integrated to develop robust classification models. This is complemented by dynamic distribution alignment, which refines the classifier iteratively. Additionally, the ensemble learning module aggregates the classifiers obtained at different stages of the optimization process, which leverages the diversity of the classifiers to enhance the overall prediction accuracy. The proposed M3D framework is evaluated on three benchmark EEG emotion recognition datasets using two validation protocols (cross-subject single-session and cross-subject cross-session), as well as on a clinical EEG dataset of Major Depressive Disorder (MDD). Experimental results demonstrate that M3D outperforms traditional non-deep learning methods, achieving an average improvement of 6.67\%, while achieving deep learning-comparable performance with significantly lower data and computational requirements. These findings highlight the potential of M3D to enhance the practicality and applicability of aBCIs in real-world scenarios.
\end{abstract}


\section{Introduction}
\label{sec:introduction}
\IEEEPARstart{E}{motion}, functioning as both a physiological and psychological state within an individual, serves as a rich repository of insights into mental and physical well-being \cite{2001Toward,2006The,Kraynak2019Neural}. Emotion recognition stands as a crucial method for objectively discerning human emotional states. Currently, the practice relies on the analysis of either non-physiological signals or physiological signals. In contrast to cues like facial expressions, voice intonation, and gestures, Electroencephalography (EEG) stands out as a non-invasive brain information collection technique \cite{liang2021eegfusenet,ye2025semi}. Its strong correlation with human emotional states, coupled with its resistance to deception, positions EEG as a highly accurate means to objectively reflect human emotions \cite{1997Feature,teplan2002fundamentals,alarcao2017emotions}.

EEG signals exhibit non-stationary characteristics and individual differences \cite{2015Transfer,si2023cross}, leading to significant variations in the distribution of EEG signals within the same subject across different time periods or between different individuals. Traditional machine learning methods assume independence and identical distribution of data, causing emotion recognition models to have poor generalization across different sessions and different individuals. This limitation poses a challenge in expanding the widespread adoption and application of these models to new sessions and individuals. On the other hand, EEG signals encounter the challenge of a small sample size, typically comprising only a few hundred to a few thousand samples \cite{liang2019unsupervised}. The labeling process is often expensive and time-consuming, making it challenging to amass a large volume of data. This difficulty in obtaining substantial data poses a hurdle in training a reliable model.

\begin{figure}[]
\begin{center}
\includegraphics[width=0.5\textwidth]{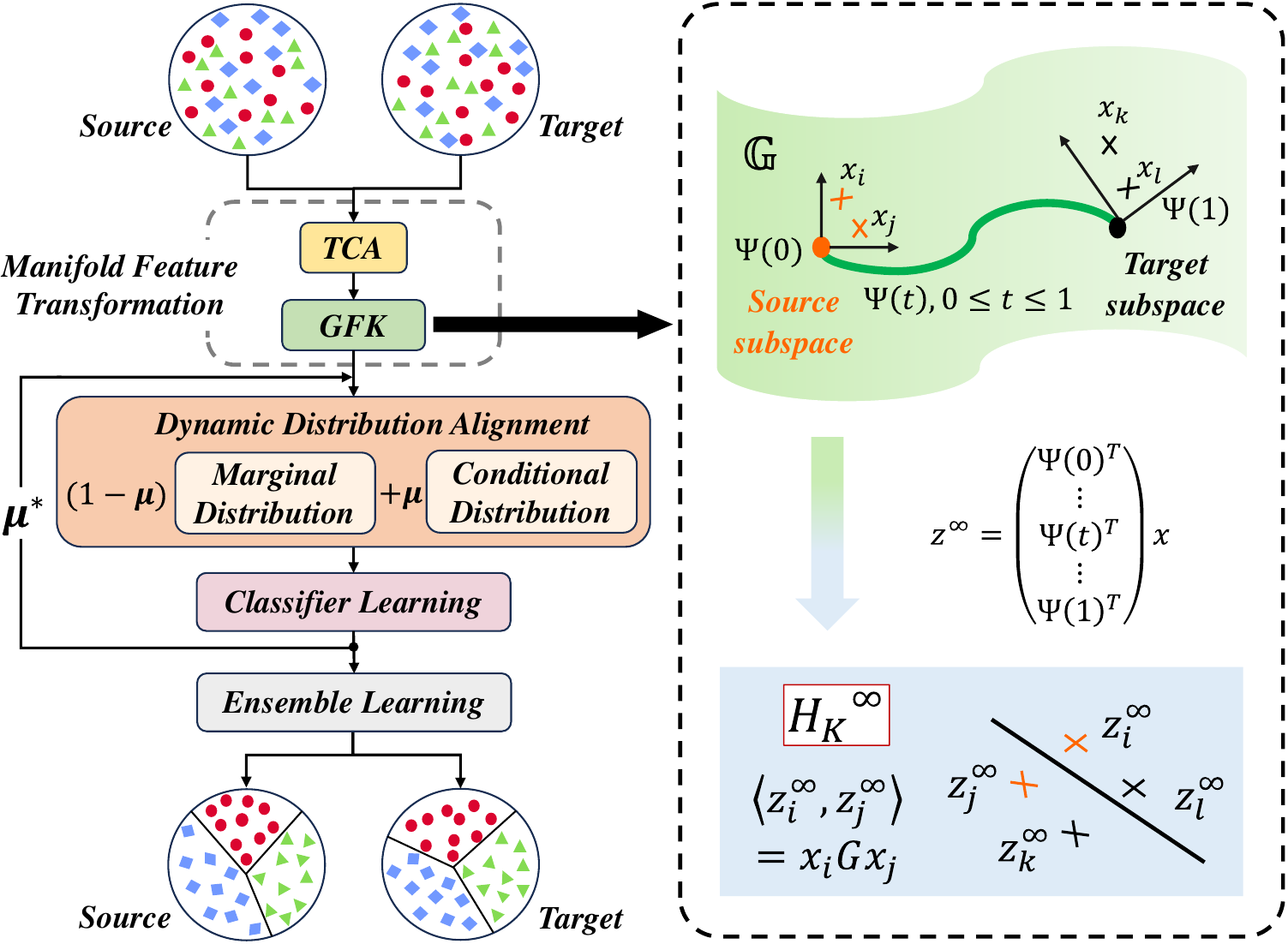}
\caption{The overall architecture of the proposed M3D model, which includes four mains modules. (1) Manifold Feature Transformation, (2) Dynamic Distribution Alignment, (3) Classifier Learning, (4) Ensemble Learning. Here, $Source$ and $Target$ refer to the source and target data, respectively. $\mathbb{G}$ is the Grassmann manifold, projected by the geodesic function $\Psi(t)$ ($t \in [0,1]$). The points $\Psi(0)$ and $\Psi(1)$ correspond to the embedded representations of the source and target domains in $\mathbb{G}$. $x_{i,j}$ and $x_{k,l}$ are the samples from the source and target data,respectively. $z^\infty_{i,j,k,l}$ are the corresponding data representation in the manifold space by the geodesic transformation. $H^\infty_{K}$ denotes the reproducing kernel Hilbert space (RKHS), and $G$ is the computed geodesics kernel.}


\label{fig:standard_pipeline}
\end{center}
\end{figure}

Early EEG-based emotion recognition methods mainly rely on conventional non-deep machine learning algorithms\cite{2015Investigating}, including Support Vector Machine (SVM), K-Nearest Neighbor (KNN), and Linear Discriminant Analysis (LDA). In these studies, there is an underlying assumption that EEG samples are independent and identically distributed. However, variations among individuals significantly undermine the effectiveness of traditional machine learning methods in cross-subject emotion recognition tasks. This results in suboptimal performance and poor generalization effects. On the other hand, transfer learning techniques can be beneficial in transferring informative data from the source (training data) to the target domain (test data), which could reduce the variations between different distributions in the modeling. More related transfer learning methods are detailed in Section \ref{sec:RelatedWork}.

Compared to deep learning-based transfer learning methods, non-deep learning-based transfer learning offers a more lightweight approach. Unlike deep learning methods, non-deep learning-based transfer requires less labeled data, which is better suited to the current conditions in EEG tasks. In this paper, we propose a novel non-deep learning-based transfer learning framework for cross-subject emotion recognition, named as Manifold-based Domain adaptation with Dynamic Distribution (M3D). The M3D model maps EEG data onto a manifold space, where both marginal and conditional distributions are dynamically analyzed and aligned. This process is coupled with classifier learning to minimize distributional differences between source and target domains. Subsequently, ensemble learning integrates the results to enhance the robustness of cross-subject and cross-session emotion recognition. Experimental results demonstrate that the proposed M3D model achieves performance comparable to existing deep learning-based methods. The main contributions of the paper are summarized below. First, a novel non-deep learning-based transfer learning model ((M3D) is proposed to address the variations in the distribution of EEG signals across different sessions and subjects. Second, both marginal distribution and conditional distribution are considered during domain alignment, where a dynamically adjusted weighting of the importance of the two distributions is incorporated. Third, an adaptive classifier learning together with an ensemble learning is introduced to iteratively refine the classifier learning process and leverage the strength of multiple learning models to optimize the prediction of data labels. Last, extensive experiments are validated on three well-known databases, as well as a clinical EEG database, with an average accuracy improvement of 6.67\%.

\section{RELATED WORK}
\label{sec:RelatedWork}
Domain adaptation targets the reduction of distribution disparities between source and target domains, achieving the assumption that the source and target data is independently and identically distributed \cite{zhou2024eegmatch}. Existing methods can be categorized into deep transfer learning and non-deep transfer learning.

\subsection{Deep Transfer Learning Methods}
The current trend in the development of transfer learning methods has mainly focused  on deep learning-based approaches. For example, Li \etal \cite{li2018cross} pioneered the use of the Domain Adversarial Neural Network (DANN) in the study of EEG-based emotion recognition. The optimization objective was to establish a shared common feature representation that mitigates distribution differences between the source and target domains. Building upon the aligned feature representation achieved with DANN, Zhou \etal \cite{zhou2023pr} introduced a prototypical representation-based pairwise learning framework, aiming to further enhance cross-subject emotion recognition performance. However, the current DANN network is limited by its exclusive focus on addressing marginal distribution differences in EEG data among different individuals, neglecting joint distribution differences. Considering that individual differences in EEG arise from joint distribution differences between different EEG signals, Li \etal \cite{2019Domain} extended the DANN network by incorporating the Joint Domain Adaptation Network. This extension provides a more comprehensive approach that considers both marginal and joint distribution differences in deep transfer learning studies.

Despite the notable achievements of the deep transfer learning methods in emotion recognition tasks, there are certain limitations. Firstly, deep learning methods demand a substantial number of training samples. However, in EEG tasks, the available samples are typically limited to a few hundred or a few thousand. Insufficient samples would constrain performance, leading to overfitting and a decrease in generalization ability. Secondly, deep transfer learning algorithms heavily rely on a substantial amount of accurately labeled source data. If there is noticeable noise in the labels from the source domain, the recognition performance would significantly deteriorate \cite{2003Class}. However, acquiring a sufficient number of samples with precise label information is not only costly but also time-consuming. Thirdly, deep transfer learning methods exhibit high computational complexity, substantial computational requirements, and involve slow and complicated training processes. The reliance on high-end hardware facilities is a significant demand, posing challenges in practical application environments where such resources may not be readily available. In contrast, non-deep transfer learning models exhibit lower complexity, reduced hardware requirements, and lightweight properties, showing the potential for good performance in widespread applications in real-life scenarios.

\subsection{Non-Deep Transfer Learning Methods}
Non-deep transfer learning methods align the feature distribution between source and target domains using conventional machine learning approaches. For example, Pan \etal \cite{5640675} introduced a Transfer Component Analysis (TCA) algorithm to learn transfer information via reproducing kernel Hilbert space and mitigate the marginal distribution differences by maximizing the Mean Discrepancy (MMD). Zheng \etal \cite{zheng2016personalizing} introduced a Transductive Parameter Transfer (TPT) algorithm, which maps personalized classifiers parameters  to the target domain while minimizing the MMD for marginal distribution. Wang \etal \cite{2018Stratified} introduced a Stratified Transfer Learning (STL) algorithm to construct a weak classifier with source data and explore intra-class relationships for adaptive spatial dimensionality reduction.

Although existing transfer learning strategies consider marginal or conditional distributions to reduce distribution differences, actual EEG disparities between subjects are in the joint distribution \cite{2019Domain}. Current algorithms focus on either marginal or conditional distributions, leading to inefficient data use. He \cite{2019Domain} proposed Joint Distribution Adaptation (JDA) to match marginal and conditional distributions equally for joint distribution adjustment \cite{2019Fault}. However, existing methods assume equal contributions from conditional and marginal probability distributions, which is unrealistic. The importance of these distributions in the joint probability distribution varies based on domain similarity. Aligning marginal distribution is crucial when domains are highly dissimilar, while prioritizing conditional distribution is beneficial when domains are similar. Dynamically adjusting the weighting of marginal and conditional distributions based on data characteristics is a key issue in non-deep transfer learning.

Most existing approaches align source and target distributions directly in the original feature space, which can be ineffective due to domain shifts. Some methods leverage the Grassmann manifold for subspace learning to transform data and extract domain-invariant features \cite{baktashmotlagh2013unsupervised, SA2013, wang2018visual}, highlighting the need for feature transformation before distribution alignment.


To address these challenges and improve non-deep transfer learning for EEG-based emotion recognition, we propose the M3D model. It first uses TCA on differential entropy features from raw EEG signals to reduce dimensionality and narrow the distribution gap. Then, it embeds features into the Grassmann manifold via Geodesic Flow Kernel (GFK) \cite{gong2012geodesic}. An adaptive factor is introduced to assess the significance of marginal and conditional distributions for better embedding. Upon achieving a robust manifold feature representation, we proceed with classifier learning to optimize the classification performance under the complex conditions of cross-subject and cross-session scenarios. In this phase, conventional machine learning methods like SVM, KNN, etc., are used as initial classifiers. Finally, an ensemble learning approach harmonizes classification results, enhancing the robustness and reliability of the framework for more accurate outcomes.

\section{METHODOLOGY}
\label{sec:Methodology}
The proposed M3D model, shown in Fig. \ref{fig:standard_pipeline}, includes four main modules: manifold feature transformation, dynamic distribution alignment, classifier learning, and ensemble learning. First, input data is reduced to lower-dimensional features using TCA, and a manifold kernel $G$ is learned to map the features into an optimal manifold space. Then, dynamic distribution alignment uses Structural Risk Minimization (SRM) \cite{Vapnik1999An} to adaptively align feature distributions within this space, improving adaptability to data variations. Next, a classifier is built on the aligned features, initially trained with traditional methods and iteratively refined by optimizing classification loss. Finally, ensemble learning combines all optimized classifiers from the iterations to produce more robust results for cross-subject and cross-session EEG-based emotion recognition.

\subsection{Manifold Feature Transformation}

Traditional feature alignment methods typically rely on original data features, which can be problematic due to deformations within the original feature space. These deformations make it difficult to effectively reduce differences between the source and target domains, negatively impacting model performance. Alternatively, the manifold space offers a solution by capturing the core aspects of data and representing original information in a more compact form \cite{2006Manifold,hamm2008grassmann}. Importantly, features embedded in the manifold space often exhibit favorable geometric properties, enhancing their effectiveness in domain adaptation \cite{baktashmotlagh2013unsupervised,SA2013,wang2018visual}. In the present study, our emphasis centers on leveraging the Grassmann manifold space to encapsulate the intrinsic essence of the data, which enables the creation of a more robust and dimensionality-reduced representation and offers a more effective way to align features across different domains.

In this paper, we introduce a manifold feature transformation process, as illustrated in Fig. \ref{fig:standard_pipeline}. The source and target domains are represented as $D_{S}=\{(x_1,y_1),...,(x_n,y_n)\}$ and $D_{T}=\{(x_{n+1},y_{n+1}),...,(x_{n+m},y_{n+m})\}$, respectively, with each domain featuring D-dimensional data representations.

Through TCA\cite{5640675}, we map both domains into a lower $d$-dimensional feature space ($d \ll D$), denoted as $T_S$ and $T_T$. This dimensionality reduction process simplifies data, facilitating more efficient data analysis and interpretation. Subsequently, we embed the $d$-dimensional feature subspace into the Grassmann manifold, denoted as $\mathbb{G}$. Our goal in this manifold space is two-fold. We aim to increase the similarity between the source and target domains' data. Simultaneously, we want to retain the unique attributes and inherent characteristics of each domain’s data. This approach ensures that while aligning the domains in a shared feature space, their distinctiveness remains intact.

To make the data distributions of the two domains more similar in the Grassmann manifold, we construct geodesics within $\mathbb{G}$, treating the source and target domains as points in this manifold space. By moving along these geodesics, we can gradually align the data distributions of the two domains. When constructing geodesics, we apply the standard Euclidean metric to the Riemannian manifold. This allows us to describe the geodesic using a parameterization function $\Psi(t)$, where $t$ ranges from 0 to 1. Specifically, $\Psi(0)$ and $\Psi(1)$ correspond to the embedded representations of the source and target domains in $\mathbb{G}$, respectively. The geodesic is the path from $\Psi(0)$ to $\Psi(1)$, representing the most direct connection between the two domain representations in the manifold space. To ensure a smooth and continuous transformation across the manifold, the intermediate values of $t$ ($t \in (0,1)$), $\Psi(t)$ satisfies the following conditions:
\begin{equation}
\label{Eq:geodesic}
\Psi\left(t\right)=T_{S}U_{1}\Gamma \left(t\right)-R_{S}U_{2}\Sigma \left(t\right),
\end{equation}
where $T_{S} \in \mathbb{R}^{M \times d}$ denote the set of orthonormal base for the source domain. $M$ is the sample size. $R_{S} \in \mathbb{R}^{M \times (D-d)}$ signifies the orthogonal complement of $T_{S}$, serving to encapsulate the dimensions outside the primary subspace spanned by $T_{S}$. $U_{1} \in \mathbb{R}^{d \times d}$ and $U_{2} \in \mathbb{R}^{(D-d) \times d}$ are both orthogonal matrices, which can be obtained through the following two sets of Singular Value Decomposition (SVD) as
\begin{equation}
\label{Eq:SVD}
T^{\top}_{S}T_{T}=U_{1}\Gamma V^{\top}, R^{\top}_{S}T_{T}=-U_{2}\Sigma V^{\top}.
\end{equation}
where $T_{T} \in \mathbb{R}^{M \times d}$ is the orthonormal base set for the target domain. $\Gamma$ and $\Sigma$ are $d \times d$ diagonal matrices. Their diagonal elements are $\cos{\theta_{k}}$ and $\sin{\theta_{k}}$ ($k = 1,2,...,d$), respectively. The variable $\theta_{k}$ represents the angles between the orthonormal bases $T_{S}$ and $T_{T}$, with $0 \leq \theta_{1} \leq \theta_{2} \leq ... \leq \theta_{d} \leq \dfrac{\pi}{2}$. $\theta_{k}$ measures the alignment between the domains in $\mathbb{G}$.

The operation $\Psi(t)^{\top}x$ projects a feature vector $x$ from the $d$-dimensional feature space onto the Grassmann manifold space $\mathbb{G}$, where $x$ could be any data sample from the source or target domain. Selecting an appropriate $t$ value is crucial, as it directly influences the projection quality of $x$. The optimal $t$ balances domain adaptation and the preservation of original features. This balance is essential for effective domain adaptation, as it allows for the integration of domain-specific features while minimizing the loss of critical information during the transition.

We introduce an integration strategy to find the optimal $t$ or a set of $t$ points between any two data samples $x_{i}$ and $x_{j}$, and utilize a geodesic kernel to facilitate the mapping process. This method transforms the $d$-dimensional feature space into an infinite-dimensional feature space, reducing domain drift. For each pair of data samples $x_{i}$ and $x_{j}$, we iteratively calculate their projections onto $\Psi(t)$ as $t$ ranges from 0 to 1. Each data sample is mapped across the Grassmann manifold space, capturing the evolution of its feature representation during the domain transition. The result of this process is two infinite-dimensional feature vectors, denoted as $z^\infty_{i}$ and $z^\infty_{j}$,
\begin{equation}
\label{Eq:i_feature}
z^\infty_{i}=\int^1_{0}\left(\Psi\left(t\right)^{\top}x_{i}\right)dt,
\end{equation}
\begin{equation}
\label{Eq:j_feature}
z^\infty_{j}=\int^1_{0}\left(\Psi\left(t\right)^{\top}x_{j}\right)dt,
\end{equation}
where $z^\infty_{i}$ and $z^\infty_{j}$ represent the entire trajectory of each data sample in the manifold, providing a comprehensive view of the domain adaptation process. This strategy improves the adaptability and accuracy of domain adaptation. From \cite{gong2012geodesic}, the geodesic kernel is the inner product of the two infinite-dimensional vectors, as
\begin{equation}
\label{Eq:geodesic_kernel}
\begin{split}
\left<z^\infty_{i},z^\infty_{j}\right>&=\int^1_{0}\left(\Psi\left(t\right)^{\top}x_{i}\right)^{\top}\left(\Psi\left(t\right)^{\top}x_{j}\right)dt \\
&=x^{\top}_{i}Gx_{j},
\end{split}
\end{equation}
where $G \in \mathbb{R}^{D \times D}$ is a positive semi-definite matrix, calculated as
\begin{equation}
\label{Eq:G}
G=\begin{bmatrix}T_{S}U_{1} & R_{S}U_{2}\end{bmatrix} \begin{bmatrix}\Lambda_{1} & \Lambda_{2} \\ \Lambda_{2} & \Lambda_{3} \end{bmatrix} \begin{bmatrix}
U^{\top}_{1} & T^{\top}_{S} \\U^{\top}_{2} & R^{\top}_{S} \end{bmatrix}.
\end{equation}
Here, $\Lambda_{1}, \Lambda_{2}, \Lambda_{3}$ are diagonal matrices, and the corresponding diagonal elements are given as
\begin{equation}
\label{Eq:lambad}
\begin{split}
\begin{cases}
    \lambda_{1,k}=1+\dfrac{\sin{2\theta_{k}}}{2\theta_{k}},\\
    \lambda_{2,k}=\dfrac{\cos{2\theta_{k}}-1}{2\theta_{k}},\\
    \lambda_{3,k}=1-\dfrac{\sin{2\theta_{k}}}{2\theta_{k}},\\
\end{cases}
\end{split}
\end{equation}
where $k \in [1,d]$. The computed geodesics kernel $G$ enables the effective transformation of a data sample $x$, initially positioned within the $d$-dimensional feature space, into the Grassmann manifold space, as
\begin{equation}
\label{Eq:manifold feature z}
z=\sqrt{G}x.
\end{equation}
This transformed representation, $z$, integrates the intrinsic features of $x$ as it is projected through the manifold, leveraging the geometrical properties of $\mathbb{G}$ to achieve a more domain-adaptive feature representation and address domain drift in transfer learning. In the implement, we use DE features for manifold feature transformation across all databases.

\subsection{Dynamic Distribution Alignment}
After the manifold feature transformation, the feature distributions between the source and target domains would exhibit greater similarity than seen with the original feature representation. Still, differences remain in both the marginal probability distribution and conditional probability distribution. Current distribution adaptation methods commonly assume equal importance for both {the marginal distribution $P$ and conditional distribution $Q$ \cite{2013Transfer,2017Joint}}. This assumption, however, may not always be applicable. In scenarios where significant differences exist between the source and target domains, the adaptation of the marginal distribution takes on increased importance. Conversely, in situations where the source and target domains are more closely aligned, the adaptation of the conditional distribution becomes more critical. Therefore, recognizing and adjusting the emphasis on marginal or conditional distribution adaptation based on the domains' specific characteristics is essential for achieving successful distribution adaptation.

In the present study, an adaptive factor $\mu \in [0,1]$ is incorporated to dynamically adjust the emphasis placed on marginal and conditional distributions. This adaptive factor allows for the flexible allocation of importance between the two types of distributions based on the characteristics of the source and target domains, which is defined as
\begin{equation}
    \label{Eq:dynamic distribution adaptation}
    \begin{split}
    \overline{D}_{f}\left(T_{S},T_{T}\right)=\left(1-\mu\right)D_{f}\left(P_{s},P_{t}\right) \\
    +\mu\sum^C_{c=1}D_{f}^{(c)}\left(Q_{s},Q_{t}\right),    
    \end{split}
\end{equation}
where $\overline{D}_{f}\left(T_{S}, T_{T}\right)$ represents the dynamic distribution adaptation between the source and target domains within the manifold feature space. $D_{f}(P_{s}, P_{t})$ is the marginal distribution adaptation, and $D_{f}^{(c)}(Q_{s}, Q_{t})$ is the conditional distribution adaptation for the class $c$ ($c \in \{1,..., C\}$). The total number of classes is represented by $C$. When $\mu$ is close to 0, it indicates significant differences between the source and target domains, thus prioritizing the adaptation of marginal distributions. When $\mu$ approaches 1, it suggests a greater similarity between the domains, thereby emphasizing the need for conditional distribution adaptation. At the midpoint, $\mu = 0.5$, the model treats both marginal and conditional distributions with equal importance.

In (\ref{Eq:dynamic distribution adaptation}), the marginal distribution adaptation $D_{f}(P_{s}, P_{t})$ is defined as
\begin{equation}
    \label{Eq:marginal distribution adaptation}
    D_{f}\left(P_{s},P_{t}\right)=\Vert E\left[f\left(z_{s}\right)\right]-E\left[f\left(z_{t}\right)\right] \Vert^2_{H_{K}},
\end{equation}
where $z_{s}$ and $z_{t}$ are the source and target data represented in the manifold feature space, given in (\ref{Eq:manifold feature z}). $f$ refers to the classifier. The term $H_{K}$ denotes the reproducing kernel Hilbert space (RKHS), which is a space of functions generated by the feature mapping $\Psi(\cdot)$. This mapping is critical for capturing the complex structures within the data by projecting it into a higher-dimensional space where linear separability is more feasible. For the computation of the conditional distribution adaptation $D_{f}^{(c)}(Q_{s}, Q_{t})$, it calculate in a similar way, as
   \begin{equation}
    \label{Eq:conditional distribution adaptation}
    D_{f}^{(c)}\left(Q_{s},Q_{t}\right)=\Vert E\left[f\left(z_{s}^{(c)}\right)\right]-E\left[f\left(z_{t}^{(c)}\right)\right] \Vert^2_{H_{K}},
\end{equation}
{where }$z_{s}^{(c)}$ and $z_{t}^{(c)}$ are the $c$-class source and target data represented in the manifold feature space (\ref{Eq:manifold feature z}). Then, (\ref{Eq:dynamic distribution adaptation}) could be rewritten as
\begin{equation}
    \label{Eq:renew dynamic distribution adaptation}
    \begin{split}
     \overline{D}_{f}\left(T_{S},T_{T}\right)=\left(1-\mu\right)\Vert E\left[f\left(z_{s}\right)\right]-E\left[f\left(z_{t}\right)\right] \Vert^2_{H_{K}} \\
     +\mu\sum^C_{c=1}\Vert E\left[f\left(z_{s}^{(c)}\right)\right]-E\left[f\left(z_{t}^{(c)}\right)\right] \Vert^2_{H_{K}}.
    \end{split}
\end{equation}

Given the absence of label information for the target domain during the training phase, we cannot directly obtain $f(z_{t}^{(c)})$ in (\ref{Eq:renew dynamic distribution adaptation}). To tackle this issue, we approximate $f(z_{t}^{(c)})$ using {the class-conditional probability distribution \cite{2017Balanced}}. Specifically, the process begins by training an initial weak classifier, such as a SVM or KNN, on the source data. This initial classifier, $f$, is subsequently utilized to infer pseudo-labels for the target data, serving as a provisional substitute for the unavailable true labels of the target domain. To enhance the reliability and accuracy of the pseudo-labels, an iterative refinement strategy is employed to improve the decodability $f$. This involves using the outcomes of the initial classifier to iteratively train subsequent classifiers, thereby progressively refining the pseudo-labels. This iterative process aims to converge towards more accurate pseudo-labels, thus improving the model's ability to generalize from the source to the target domain effectively. Further details on the specific iterative refinement process and how it contributes to classifier learning will be discussed in Section \ref{subsec:Classifier Learning}.

In (\ref{Eq:renew dynamic distribution adaptation}), the adaptive factor $\mu$ is not a constant value predetermined by prior knowledge. Instead, it is dynamically learned based on the underlying data distribution. This parameter comprehensively incorporates both the marginal distribution difference and conditional distribution difference between domains. Specifically, we use {the A-distance measurement \cite{ben2006analysis}} as a metric to estimate the marginal distribution difference, denoted as $d_{A}$, between the source and target domains, as
\begin{equation}
    \label{Eq:a-distance in marginal distribution}
    d_{A}\left(T_{S},T_{T}\right)=2\left(1-2\epsilon\left(h\right)\right),
\end{equation}
where $\epsilon\left(h\right)$ denotes the hinge loss from a binary classifier distinguishing between samples from the source and target domains. For the conditional distribution difference, based on the provided label information for source data and the estimated pseudo-label information of target data, we conduct the A-distance measurement within each $c$ class ($c \in [1,\dots, C]$) as
\begin{equation}
    \label{Eq:a-distance in conditional distribution}
    d_{c}=d_{A}\left(T_{S}^{(c)},T_{T}^{(c)}\right).
\end{equation}
Here, $T_{S}^{(c)}$ and $T_{T}^{(c)}$ represent the subsets of data corresponding to class $c$ from the source and target domains, respectively. Then, the adaptive factor $\mu$ can be estimated as
\begin{equation}
    \label{Eq:mu}
    \mu=1-\dfrac{d_{A}}{d_{A}+\sum^C_{c=1}d_{c}}.
\end{equation}
In the process of dynamic distribution alignment, the adaptive factor $\mu$ is recalculated at each iteration, underscoring the iterative and responsive nature of this adaptation strategy. This re-calibration is essential to ensure that the adaptation process remains attuned to the evolving similarities and differences between the source and target domain feature distributions as they are represented in the manifold feature space. More detailed descriptions of $\mu$ are presented in the Supplementary Materials Appendix A.

\subsection{Classifier Learning}

\label{subsec:Classifier Learning}
Through the integration of manifold feature learning and dynamic distribution alignment, we can formulate an adaptive classifier $f$ at the $\iota$th iteration as
\begin{equation}
\label{Eq:f}
\begin{split}
    f^{\iota}=\mathop{\arg\min}_{f\in H_{K}}\sum^n_{i=1}\left(y_{i}-f^{(\iota-1)}\left(z_{i}\right)\right)^2+\eta\Vert f^{(\iota-1)}\Vert^2_{H_{K}} \\
    +\lambda\overline{D}_{f^{(\iota-1)}}\left(T_{S},T_{T}\right)
    +\rho R_{f^{(\iota-1)}}\left(T_{S},T_{T}\right).
\end{split}
\end{equation}
Here, $\iota$ is the iteration number, $y_i$ is the groundtruth, and $f(z_i)$ is the predicted classification result. $\Vert \cdot \Vert^2_{H_{K}}$ refers to the L2 norm in the reproducing kernel Hilbert space. The dynamic distribution adaptation term, $\overline{D}_{f}\left(T_{S},T_{T}\right)$, is obtained in (\ref{Eq:renew dynamic distribution adaptation}). Further, a Laplacian regularization term, $R_{f}\left(T_{S},T_{T}\right)$, is incorporated to leverage the inherent geometric similarities among neighboring points {in the manifold $\mathbb{G}$ \cite{2006Manifold}}. Here, the classifier $f$ can be initialized using a commonly employed machine learning classifier, followed by adaptive optimization through an iterative learning process. $\eta, \lambda$ and $\rho$ are regularization parameters.

In the SRM framework with {the representer theorem \cite{2006Manifold}}, $f$ can be expanded as
    \begin{equation}
    \label{Eq:SRM f} f\left(z\right)=\sum^{n+m}_{i=1}\beta_{i}K\left(z_{i},z\right),
    \end{equation}
which includes both labeled and unlabeled samples from source and target domains, and  $\beta = (\beta_{1}, \beta_{2}, \dots, \beta_{n+m})^T \in \mathbb{R}^{(n+m)\times 1}$ is the coefficient vector. $K$ is the kernel function. For the SRM on the source domain, the first part in (\ref{Eq:f}) could be expressed as  
    \begin{equation}
        \label{Eq:SRM on the sourece domain}
        \begin{split}
        \sum^n_{i=1}\left(y_{i}-f\left(z_{i}\right)\right)^2+\eta\Vert f\Vert^2_{H_{K}}\\
        =\sum^{n+m}_{i=1}A_{ii}\left(y_{i}-f\left(z_{i}\right)\right)^2+\eta\Vert f\Vert^2_{H_{K}}, \\
        \end{split}
     \end{equation}
where $\Vert \cdot\Vert_{F}$ is the Frobenius norm. The matrix $A \in \mathbb{R}^{(n+m) \times (n+m)}$ is {a diagonal indicator matrix with $A_{ii} = 1$ if $i \in T_{S}$, and $A_{ii} = 0$ if $i \in T_{T}$}. Substituting (\ref{Eq:SRM f}) into (\ref{Eq:SRM on the sourece domain}), then we can obtain
\begin{equation}
\label{final SRM on the sourece domain}
\begin{split}
    \sum^n_{i=1}\left(y_{i}-f\left(z_{i}\right)\right)^2+\eta\Vert f\Vert^2_{H_{K}}\\
    =\Vert\left(Y-\beta^{\top}\boldsymbol{K}\right)A\Vert^2_{F}+\eta tr\left(\beta^{\top}\boldsymbol{K}\beta\right),\\
\end{split}
\end{equation}
where $\boldsymbol{K} \in \mathbb{R}^{(n+m) \times (n+m)}$ is the kernel matrix, with $\boldsymbol{K}_{ij} = \boldsymbol{K}(z_{i},z_{j})$. The label information of the source and target data is denoted by $Y = [y_{1},...,y_{n+m}]$. For the target data, the corresponding label information is the pseudo label generated by the learned classifier $f(\cdot)$. $tr(\cdot)$ is the trace operation.

For the second part in (\ref{Eq:f}), through applying (\ref{Eq:SRM f}), we can represent the calculation of dynamic distribution adaptation (\ref{Eq:renew dynamic distribution adaptation}) as
    \begin{equation}
    \label{Eq:final dynamic distribution adaptation}
    \overline{D}_{f}\left(T_{S},T_{T}\right)   =tr\left(\beta^{\top}\boldsymbol{K}M\boldsymbol{K}\beta\right),
    \end{equation}
where $M = (1-\mu)M_{0} + \mu\sum^C_{c=1}M_{c}$. $c$ is the class category, and $C$ refers to the total number of classes. $M_{0}$ and $M_{c}$ are defined as
\begin{equation}
    \label{Eq:M0}
     \left(M_{0}\right)_{ij}=  
     \begin{cases}
     \frac{1}{n^2}& z_{i},z_{j}\in T_{S} \\
     \frac{1}{m^2}& z_{i},z_{j}\in T_{T} \\
     -\frac{1}{mn}& otherwise
     \end{cases},
\end{equation}
    \begin{equation}
    \label{Eq:Mc}
    \left(M_{c}\right)_{ij}=  
    \begin{cases}
     \frac{1}{n^2_{c}}& z_{i},z_{j}\in T_{S}^{(c)}\\
     \frac{1}{m^2_{c}}& z_{i},z_{j}\in T_{T}^{(c)} \\
     -\frac{1}{m_{c}n_{c}}& 
     \begin{cases}
     z_{i}\in T_{S}^{(c)},z_{j}\in T_{T}^{(c)} \\
     z_{i}\in T_{T}^{(c)},z_{j}\in T_{S}^{(c)}
     \end{cases}\\
     0 & otherwise 
     \end{cases}.
\end{equation}
Here, $n_{c} = |T_{S}^{(c)}|$ and $m_{c} = |T_{T}^{(c)}|$. The third part in (\ref{Eq:f}) (the Laplacian regularization term $R_{f}\left(T_{S},T_{T}\right)$) could be computed as
    \begin{equation}
    \label{Eq:Laplacian regularization}
    \begin{split}    R_{f}\left(T_{S},T_{T}\right)&=\sum^{n+m}_{i,j=1}W_{ij}\left(f\left(z_{i}\right)-f\left(z_{j}\right)\right)^2 \\
        &=\sum^{n+m}_{i,j=1}f\left(z_{i}\right)L_{ij}f\left(z_{j}\right),\\
    \end{split}                                       
    \end{equation}
where $W_{ij}$ is a pairwise similarity matrix, and $L=D-W$ is the normalized graph Laplacian matrix. $D$ is a diagonal matrix, with $D_{ii}=\sum^{n+m}_{j=1}W_{ij}$. $W$ could be represented as
    \begin{equation}
    \label{Eq:Wij}
    W_{ij}=
    \begin{cases}
    sim\left(z_{i},z_{j}\right) & z_{i}\in N_{p}\left(z_{j}\right) or z_{j}\in N_{p}\left(z_{i}\right),\\
    0& otherwise,
    \end{cases}
    \end{equation}
where $sim(\cdot,\cdot)$ is a similarity function (e.g., cosine distance) used to assess the proximity between two points. $N_{p}(z_{i})$ represents a set of $p$ nearest points to $z_{i}$. $p$ is a hyperparameter.

Similarly, by applying (\ref{Eq:SRM f}), the Laplacian regularization given in (\ref{Eq:Laplacian regularization}) can be expressed as
\begin{equation}
\label{Eq:renew Laplacian regularization}
R_{f}\left(T_{S},T_{T}\right)=tr\left(\beta^{\top}\boldsymbol{K}L\boldsymbol{K}\beta\right).
\end{equation}

Then, by substituting (\ref{final SRM on the sourece domain}), (\ref{Eq:final dynamic distribution adaptation}) and (\ref{Eq:renew Laplacian regularization})into the (\ref{Eq:f}), the classifier $f$ can be calculated as:
    \begin{equation}
    \label{Eq:final f}
    \begin{split}
    f=\mathop{\arg\min}_{f\in H_{K}}\Vert\left(Y-\beta^{\top} \boldsymbol{K}\right)A\Vert^2_{F}+\eta tr\left(\beta^{\top} \boldsymbol{K}\beta\right) \\
    +tr\left(\beta^{\top} \boldsymbol{K}\left(\lambda M+\rho L\right)\boldsymbol{K}\beta\right).
    \end{split}
    \end{equation}

Setting $\frac{\partial f}{\partial \beta} = 0$, (\ref{Eq:final f}) could be solved as:
    \begin{equation}
    \label{Eq:beta}
    \begin{split}
    \beta^{*}=\left(\left(A+\lambda M+\rho L\right)\boldsymbol{K}+\eta I \right)^{-1}AY^{\top}.   
    \end{split}
    \end{equation}  
After obtaining $\beta^{*}$, the classifier $f$ can be determined using (\ref{Eq:final f}). Then, the corresponding soft label of the target domain at $\iota$th iteration could be denoted as $\hat{y}_{t}^{\iota}=f^{\iota}(z_{t})$. In the implementation, through $l$ iterations, the adaptive factor $\mu$ in the dynamic distribution alignment is continuously updated, and the classifier learning leverages the corresponding updated $\mu$ to further refine the pseudo label of the target domain.

\begin{figure}[]
\begin{center}
\includegraphics[width=0.5\textwidth]{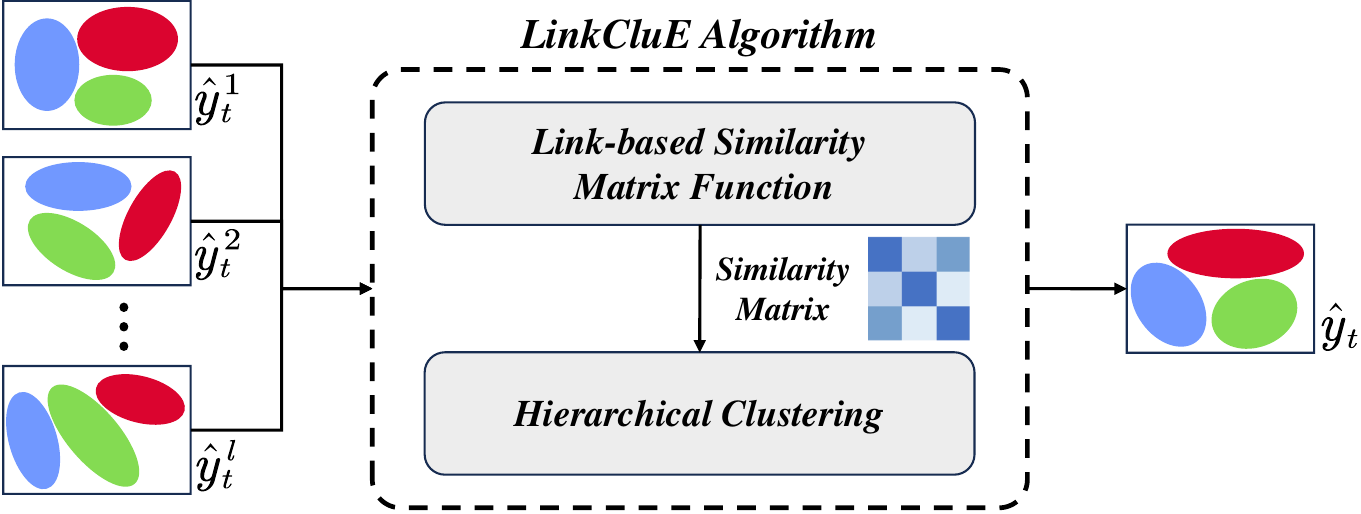}
\end{center}
\caption{The main process of the LinkCluE ensemble algorithm. (1) Link-based Similarity Matrix Function: takes a set of {input vectors $\hat{y}_{t}^{\iota}$} ($\iota=1,\dots,l$) and constructs a similarity matrix as its output. (2) Hierachical Clustering: processes the similarity matrix through a hierarchical clustering algorithm to produce the final clustering result, {denoted as $\hat{y}_{t}$}.}
\label{fig:Ensemble_Learning}
\end{figure} 

\begin{algorithm}[h]
    \caption{The algorithm flow of the proposed M3D model.}
    \label{alg::conjugateGradient}
    \begin{algorithmic}[1]
        \Require
    \renewcommand{\algorithmicrequire}{\textbf{}}
    \Require - Data matrix $X=\left[X_{s},X_{t}\right]$, source labels $y_{s}$, subspace dimension $d$, regularization parameters $\eta,\lambda,\rho$ and the iteration number $\iota$; 
    \Ensure Classifier $f(\cdot)$;
    \Statex \textcolor{gray}{\# Manifold Feature Transformation}
    \State Learn the manifold feature transformation kernel $G$ and obtain the manifold feature representation $z$ using (\ref{Eq:manifold feature z});
    \State Train a weak classifier based on the source data $D_{S}$, and get initialized soft labels $\hat{y}_{t}^{0}$ for the target data $D_{T}$;
    \State Construct the geodesic line kernel $K$ using the transformed features $z_{s}=z_{1:n}$ and $z_{t}=z_{n+1:n+m}$;
    \For {$\iota=1$ to $l$}
    \Statex \textcolor{gray}{\# Dynamic Distribution Alignment}
        \State Calculate the adaptive factor $\mu$ using (\ref{Eq:mu}), and compute $M_{0}$ and $M_{c}$ using (\ref{Eq:M0}) and (\ref{Eq:Mc});
    \Statex \textcolor{gray}{\# Classifier Learning}
        \State Solve the equation using (\ref{Eq:beta}) to compute $\beta^{*}$ and obtain the classifier $f$ using (\ref{Eq:final f});
        \State Update the soft labels for the target data $\hat{y}_{t}^{\iota}=f(z_{t})$;
    \EndFor
    \State Return the classifier $f$;
    \Statex \textcolor{gray}{\# Ensemble Learning}
         \State Perform ensemble learning based on the obtained soft labels in the loop $\{\hat{y}_{t}^{\iota}\}_{\iota=1}^l$.
    \end{algorithmic}
\end{algorithm}

\subsection{Ensemble Learning}
Ensemble learning with {LinkCluE algorithm \cite{iam2010linkclue}} is adopted here to strategically reduce classification bias inherent in individual models and improve the overall generalization performance of the model. Based on the obtained $\hat{y}_{t}$ in a certain round of classifier learning, the ensemble learning leverages the strengths over a group of $\hat{y}_{t}^{\iota}$ ($\iota=1,\dots,l$. Empirically, $l$ is set to 10) and form together to generate a strong prediction results $\hat{y}_{t}$.

Fig. \ref{fig:Ensemble_Learning} illustrates the two main steps involved in the LinkCluE algorithm: the link-based similarity matrix function and the consensus function. For the link-based similarity matrix function, it constructs a similarity matrix based on the links or relationships between data points in terms of Connected-Triple-Based Similarity (CTS), SimRank-Based Similarity (SRS), and Approximate SimRank-Based Similarity (ASRS). This matrix serves as a foundation for understanding the complex, often non-linear relationships that exist between data points, leveraging the structure of the data to enhance the learning process. For the consensus function, it integrates the outcomes of various individual models within the ensemble by leveraging three different hierarchical agglomerative clustering algorithms, Single Linkage (SL), Complete Linkage (CL), and Average Linkage (AL). This step is crucial for mitigating individual model biases and errors, leading to more accurate and generalizable results. 

The overall algorithm of the proposed M3D is illustrated in Algorithm \ref{alg::conjugateGradient}.

\begin{table}[]
\begin{center}
\caption{Cross-subject single-session LOSO CV on SEED, SEED-IV and SEED-V with different initial classifiers.}
\label{tab:seed_and_seediv and_seedv_singlesession}
\setlength{\tabcolsep}{1mm}
\scalebox{1}
{
\begin{tabular}{l|lcccccc}
\toprule
\multicolumn{2}{l}{Database} &KNN & SVM  & DT & Adaboost & GNB & Bagging   \\ 
\midrule
\multirow{7}{*}{SEED} &Accuracy & 84.46    & 84.51 & \textbf{84.57} & 84.10 &84.10 & 84.10 \\
&Sensitivity &\textbf{82.57} &80.82 & 76.04 & 81.08 & 81.08& 81.08\\
&Specificity & 85.29 &86.25 &\textbf{88.78} &85.42&85.42&85.42\\
&Precision &71.07 &73.48 &\textbf{76.94} &70.80&70.80&70.80\\
&F1-score  & 76.39 & \textbf{76.97} & 76.49 & 75.59 & 75.59& 75.59  \\
&AUROC &\textbf{88.37} &88.36 &88.36 &88.04 &88.04 &88.04 \\
&NPV &\textbf{91.79} &90.51 &88.27 &91.19 &91.19 &91.19\\
\midrule
\midrule
\multirow{7}{*}{SEED-IV} &Accuracy & 58.25    & 58.82 & \textbf{60.94} & 58.89 &58.42 & 58.25 \\
&Sensitivity &66.54 &66.05&\textbf{78.10} &65.09 &71.16 &66.54\\
&Specificity &54.72 & 55.99 &55.07 &\textbf{56.33} &54.20 &54.72\\
&Precision &\textbf{38.53} &37.10 &37.30 &38.09&33.95&\textbf{38.53}\\
&F1-score  & 48.80 & 47.51 & \textbf{50.49} & 48.06 & 45.97 &48.80  \\
&AUROC &71.65 &72.20 &\textbf{73.68}&72.32 &71.89 &71.69\\
&NPV &79.31&80.75 &\textbf{88.02 }&79.63 &85.03 &79.31\\
\midrule
\midrule
\multirow{7}{*}{SEED-V}
&Accuracy &64.39 &64.41 &\textbf{65.25} &63.05 &63.44 &63.04\\
&Sensitivity&\textbf{62.30} &\textbf{62.30} & 57.53&60.48 &\textbf{62.30}&\textbf{62.30}\\
&Specificity&64.55 &64.57&\textbf{66.16}&63.33 &63.52&63.10\\
&Precision&11.58 &11.59&\textbf{16.60}& 15.40 &11.29 &11.18\\
&F1-score&19.54&19.54&\textbf{25.76}&24.55 &19.12 &18.96\\ 
&AUROC&77.87&77.80 &\textbf{78.46} &76.92 &77.28 &77.10\\
&NPV &\textbf{95.83}&\textbf{95.83}&93.01 &93.56 &95.76 &95.74\\
\bottomrule
\end{tabular}
}
\end{center}
\end{table}

\section{Experimental Results}
\label{sec:experiment}

\subsection{Emotional Databases}

The model performance is carefully validated on three well-known publicly available databases: SEED \cite{2015Investigating}, SEED-IV \cite{Zheng2019EmotionMeter} and SEED-V\cite{liu2021comparing}. The SEED database includes EEG emotion data collected from 15 participants. Each participant was exposed to a total of 15 movie clips designed to evoke three types of emotions (positive, negative, and neutral). The SEED-IV database also includes EEG emotion data from 15 participants, exposed to a total of 24 movie clips aimed at eliciting four types of emotions (happy, sad, neutral, and fear). In the SEED-V database, a total of 5 emotions (happiness, sadness, fear, disgust, and neutral) are elicited using 15 movie clips, and the simultaneous EEG signals under the 5 emotions are recorded from 16 subjects (6 males and 10 females). For both the SEED, SEED-IV, and SEED-V databases, participants undergo three sessions spread across separate days, with a one-week break between sessions. The EEG signals are recorded using the ESI Neuroscan system with 62 channels.

The EEG preprocessing follows established procedures in the literature. This approach aligns with the standard preprocessing practices in emotion-related EEG research and ensures consistency across the widely used SEED, SEED-IV, and SEED-V datasets in the literature. The process begins with the downsampling of EEG signals to 200 Hz, followed by the removal of artifacts such as the Electrooculogram (EOG) and Electromyography (EMG). Downsampling to 200 Hz effectively reduces computational complexity while preserving the temporal resolution necessary for emotion recognition \cite{2015Investigating,Zheng2019EmotionMeter,liu2021comparing}. A bandpass filter with a 0.3-50 Hz range is then applied to improve the signal quality, as the majority of emotion-related electroencephalography activities occur below 50 Hz. Following this, each trial was segmented into multiple 1-second data samples. To extract emotional related information, DE features were extracted across five frequency bands: Delta (1-3 Hz), Theta (4-7 Hz), Alpha (8-13 Hz), Beta (14-30 Hz), and Gamma (31-50 Hz). Thus, for each 1-second data sample, a total of 310 features (5 frequency bands $\times$ 62 channels) were characterized, which serves as the input for the model. To explore EEG signal inter-subject differences in SEED and SEED-IV datasets, we perform a comprehensive statistical analysis of input EEG features via qualitative visualizations and quantitative comparisons. More detailed descriptions and results are presented in Supplementary Materials Appendix B.

To further assess the generalizability of the proposed M3D model beyond controlled experimental conditions, we extend model validation to a real-world clinical EEG database obtained from Hospital Universiti Sains Malaysia (HUSM) \cite{mumtaz2016mdd}. This dataset consists of EEG recordings from 27 healthy individuals (mean age: 38.28±15.64 years) and 29 individuals diagnosed with MDD (Major Depressive Disorder) (mean age: 40.33±12.86 years). Participant classification adhered to the international diagnostic criteria outlined in the Diagnostic and Statistical Manual-IV (DSM-IV). This dataset used different devices and settings compared to the SEED and SEED-IV databases, with EEG recordings collected using the Brain Master Discovery amplifier, 19 electrodes, and a sampling rate of 256 Hz. Preprocessing involved bandpass filtering (0.5-70 Hz) and notch filtering at 50 Hz to eliminate power-line noise. This preprocessing pipeline is widely used and validated in the literature \cite{mumtaz2016mdd} as a reliable approach for enhancing data quality and processing efficiency without compromising informative neural features. The experimental protocol included EEG recordings under both open-eye (EO) and closed-eye (EC) conditions, each lasting 5 minutes. During the EO condition, participants were instructed to remain relaxed while minimizing eye movements to maintain data quality. Similar to the SEED, SEED-IV and SEED-V databases, this database also extracts DE features across five frequency bands.

\subsection{Experimental Protocols}
In order to thoroughly assess the robustness and stability of the proposed model and ensure a detailed comparison with existing literature, two types of experimental protocols are conducted.
\begin{itemize}
    \item \textbf{Cross-subject single-session leave-one-subject-out cross-validation (Cross-subject single-session LOSO CV). The data samples from N subjects (SEED/SEED-IV: N=14, SEED-V: N=15}) in the first session are used as the source domain, and the data samples from the remaining 1 subject in the same session are used as the target domain. This procedure repeats N+1 times (SEED/SEED-IV: N=14, SEED-V: N=15), ensuring that each participant is treated as the target domain at least once.
    \item \textbf{Cross-subject cross-session leave-one-subject-out cross-validation (Cross-subject cross-session LOSO CV)}. The data samples from N subjects (SEED/SEED-IV: N=14, SEED-V: N=15) across all three sessions are used as the source domain, and the data samples from the remaining 1 subject across all three sessions are used as the target domain. This procedure also repeats N+1 times (SEED/SEED-IV: N=14, SEED-V: N=15), ensuring that each participant is treated as the target domain at least once.
\end{itemize}

\color{black}
In the present study, we adopt six widely recognized machine-learning classifiers to generate the initial classification results ($\hat{y}_t^{0}$) as mentioned in Section \ref{subsec:Classifier Learning}. The adopted machine-learning classifiers include KNN\cite{0An}, SVM\cite{1995Support}, DT\cite{2018Study}, Adaboost classifier\cite{1997A}, GNB classifier\cite{2001An}, and Bagging classifier\cite{1996Bagging}. In the SVM classifier, the radial basis function (RBF) kernel is utilized. In ensemble learning, the application of three similarity matrix functions (CTS, SRS, and ASRS) combined with three hierarchical agglomerative clustering algorithms (SL, CL, and AL) yields a total of 9 possible combinations. Here, the combination that achieves the highest accuracy in predicting target domain labels will be chosen as the optimal outcome. Notably, all the best results are achieved using the DT classifier and the CTS-SL approach. For performance evaluation, a comprehensive analysis of the model’s effectiveness is conducted using multiple metrics, including accuracy, sensitivity, specificity, precision, F1-score, Area Under the Receiver Operating Characteristic Curve (AUROC) and Negative Predictive Value (NPV) metrics. We set the manifold feature dimension $d = 128$ in all datasets except for the MDD database (where $d = 100$). The regularization parameters are set as $\eta =0.1$, $\rho = 1$ and $\lambda = 0.4$.

\begin{table}[]
\begin{center}
\caption{Performance comparison on SEED using cross-subject single-session LOSO CV. The \underline{second-best} traditional machine learning method result is underlined.}
\label{tab:seed_singleCompare}
\setlength{\tabcolsep}{2.5mm}
\scalebox{1}{
\begin{tabular}{lclc}
\toprule
Methods   & $P_{acc}$   & Methods   & $P_{acc}$    \\ 
\midrule
\multicolumn{4}{c}{\textbf{\textit{Traditional machine learning methods}}} \\ 
\midrule
SVM\cite{li2018cross} & 58.18/13.85
&TCA\cite{li2018cross} & 64.00/14.66  \\
KPCA\cite{li2018cross} &69.02/09.25
&SCA\cite{ma2019reducing} &66.33/10.60\\
DICA\cite{ma2019reducing}& 69.41/07.79
&KLIEP\cite{2018A}&45.71/17.76\\
ULSIF\cite{2018A} &51.18/13.57
&TKL\cite{2018A}       & 63.54/15.47 \\
SA\cite{2018A}        & 69.00/10.89 
&GFK\cite{2018A} & 71.31/14.09 \\
T-SVM\cite{2018A} & 72.53/14.00 
&TPT\cite{2018A}&\underline{76.31/15.89}  \\
\midrule
\multicolumn{4}{c}{\textit{\textbf{Deep learning methods}}}\\ 
\midrule
MLP\cite{li2018cross} &61.01/12.38
&DANN\cite{li2018cross}  &79.79/13.14 \\
DAN\cite{li2018cross} &83.81/08.56
& MMD \cite{2019Domain}  & 80.88/10.10  \\
ADA\cite{2019Domain}   & 84.47/10.65 
&DG-DANN\cite{ma2019reducing}&84.30/08.32 \\
DGCNN\cite{2018A} &79.95/09.02
&Bi-DANN\cite{2018A}   & 83.28/09.60\\
BiDANN-S\cite{2018A}&84.14/06.87 
&DCORAL\cite{she2023multisource}&62.14/07.98 \\
DDC\cite{she2023multisource} &74.34/09.05 
&MS-MDA\cite{she2023multisource} &79.67/08.01\\
SOGNN\cite{cheng2024emotion} &76.00/06.92
&R2G-STLT\cite{cheng2024emotion}&77.96/06.38 \\
A-LSTM\cite{li2021novel} &72.18/10.85
&TANN\cite{li2021novel}&84.41/08.75 \\
GECNN\cite{pan2023st} &82.46/10.83
&Saliency\cite{pan2023st} &84.11/02.90\\
P-GCNN\cite{pan2023st} &84.35/10.28
&ST-SCGNN\cite{pan2023st} &85.90/04.90\\
\midrule
\multicolumn{3}{l}{\textbf{M3D}} & \textbf{84.57/09.49}  \\
\bottomrule
\end{tabular}
}
\end{center}
\end{table}

\begin{table}[]
\begin{center}
\caption{Performance comparison on SEED-IV using cross-subject single-session LOSO CV. The \underline{second-best} traditional machine learning method result is underlined.}
\label{tab:seedIV_singleCompare}
\setlength{\tabcolsep}{2.8mm}
\scalebox{1}{
\begin{tabular}{lclc}
\toprule
Methods   & $P_{acc}$   & Methods   & $P_{acc}$    \\ 
\midrule
\multicolumn{4}{c}{\textbf{\textit{Traditional machine learning methods}}} \\ 
\midrule
GFK\cite{zhou2023pr}&44.04/09.31
&GFHF\cite{zhou2023pr}&49.29/07.60\\
JDA\cite{zhou2023pr}&54.57/05.47
&GAKT\cite{zhou2023pr}&58.49/06.23\\
MIDA\cite{zhou2023pr}&\underline{60.22/08.69}
&KLIEP\cite{li2021novel} &31.46/09.20\\
ULSIF\cite{li2021novel} &32.99/11.05
&STM\cite{li2021novel} &39.39/12.40\\
SVM\cite{li2021novel} &37.99/12.52
&TCA\cite{li2021novel} &56.56/13.77\\
\midrule
\multicolumn{4}{c}{\textit{\textbf{Deep learning methods}}}\\ 
\midrule
ScalingNet\cite{cheng2024emotion} &60.75/10.40
&GECNN\cite{pan2023st} &55.01/10.29\\
DANN\cite{li2021novel} &47.59/10.01
&DGCNN\cite{li2021novel} &52.82/09.23\\
A-LSTM\cite{li2021novel}&55.03/09.28
&DAN\cite{li2021novel} &58.87/08.13\\

RGNN\cite{2023MFA-LR}&53.49/10.57
&JAGP\cite{2023MFA-LR} &54.37/09.49 \\
MFA-LR\cite{2023MFA-LR} &64.96/14.04
&DDC\cite{2021MS}&37.41/06.36\\
DCORAL\cite{2021MS}&37.43/03.08
&MS-MDA\cite{2021MS}  & 59.34/05.48 \\
\midrule
\multicolumn{3}{l}{\textbf{M3D}} & \textbf{60.94/08.84}  \\
\bottomrule
\end{tabular}
}
\end{center}
\end{table}

\begin{table}[]
\begin{center}
\caption{Performance comparison on SEED-V using cross-subject single-session LOSO CV. The \underline{second-best} traditional machine learning method result is underlined. Models reproduced by the authors are indicated with an asterisk (*).}
\label{tab:seedV_singleCompare}
\setlength{\tabcolsep}{2.8mm}
\scalebox{1}{
\begin{tabular}{lclc}
\toprule
Methods   & $P_{acc}$   & Methods   & $P_{acc}$    \\ 
\midrule
\multicolumn{4}{c}{\textbf{\textit{Traditional machine learning methods}}} \\ 
\midrule
TCA*\cite{5640675} &39.32/13.18
&SA*\cite{SA2013} &33.45/09.92 \\
GFK* \cite{gong2012geodesic}&36.44/16.14
&SVM*\cite{1995Support} &49.39/18.13\\
Adaboost*\cite{1997A}  &50.01/21.71
&KNN*\cite{coomans1982alternative} &33.14/13.62 \\
KPCA*\cite{mika1998kernel} &34.41/19.21 
&CORAL*\cite{2015Return} &\underline{51.10/21.12} \\      
\midrule
\multicolumn{4}{c}{\textit{\textbf{Deep learning methods}}}\\ 
\midrule
DDC\cite{zhu2024instance} &28.70/09.40
&JDA\cite{zhu2024instance}  &29.74/06.90\\
DAN\cite{zhu2024instance}  &41.24/12.12
&JD-IRT\cite{zhu2024instance}  &60.17/11.06\\
\midrule
\multicolumn{3}{l}{\textbf{M3D}} & \textbf{65.25/07.87}  \\
\bottomrule
\end{tabular}
}
\end{center}
\end{table}

\begin{table}[]
\begin{center}
\caption{Cross-subject cross-session LOSO CV on SEED, SEED-IV and SEED-V with different initial classifiers}
\label{tab:seed_and_seediv_and_seedv_crosssession}
\setlength{\tabcolsep}{1mm}
\scalebox{1}
{
\begin{tabular}{l|lcccccc}
\toprule
\multicolumn{2}{l}{Database} &KNN & SVM  & DT & Adaboost & GNB & Bagging   \\ 
\midrule
\multirow{7}{*}{SEED} &Accuracy & 74.79    & 75.98 & \textbf{77.05} & 76.60 &76.44 & 76.47 \\
&Sensitivity &72.46 &71.15 &\textbf{72.93} &72.37 &72.00 &72,07\\
&Specificity &75.66 &78.14 &\textbf{78.92 }&78.46 &78.41 &78.42\\
&Precision &52.48 &59.28 &\textbf{61.05} &59.70&59.70&59.70\\
&F1-score  & 60.88 & 64.67 & \textbf{66.46} & 65.43 & 65.28 &65.30  \\
&AUROC &60.88&64.67 &\textbf{66.46} &65.43 &65.28 &65.30\\
&NPV &\textbf{88.10} &85.83 &86.55&86.56 &86.31 &86.36\\ 
\midrule
\midrule
\multirow{7}{*}{SEED-IV} &Accuracy & 64.55   & 64.55 & \textbf{64.90} &64.71&64.71 &64.71 \\
&Sensitivity &68.02 &68.02 &70.36&\textbf{74.10} &68.02&68.02\\
&Specificity &63.10 &63.10 &62.85 &61.23 &\textbf{63.33}&\textbf{63.33}\\
&Precision &43.53&43.53 &41.49 &41.50&\textbf{43.68}&\textbf{43.68}\\
&F1-score  &53.09 & 53.09 & 52.20 & \textbf{53.20} & \textbf{53.20} & \textbf{53.20}  \\
&AUROC &76.24&76.24&\textbf{76.46} &76.43&76.35&76.35\\
&NPV &82.51 &82.51 &84.99 &\textbf{86.43} &82.57 &82.57\\
\midrule
\midrule
\multirow{7}{*}{SEED-V}
&Accuracy &56.37 &56.61 &\textbf{59.15} &56.61 &56.36 &56.37\\
&Sensitivity &54.97 &54.97 &\textbf{58.65} &54.70&54.70&54.70\\
&Specificity &56.64 &56.93 &\textbf{59.24}&56.98 &56.69 &56.70\\
&Precision &19.85 &19.96 &\textbf{20.60} &20.00 &19.89 &19.89\\
&F1-score &29.17 &29.29 &\textbf{30.49} &29.29 &29.17 &29.17\\
&AUROC &72.76 &72.92 &\textbf{74.70} &72.92 &72.77 &72.76\\
&NPV &86.55 &86.61 &\textbf{88.82} &86.48 &86.42 &86.42\\
\bottomrule
\end{tabular}
}
\end{center}
\end{table}

\begin{table}[]
\begin{center}
\caption{Performance comparison on SEED using cross-subject cross-session LOSO CV. The \underline{second-best} traditional machine learning method result is underlined.}
\label{tab:seed_crossCompare}
\setlength{\tabcolsep}{1.8mm}
\scalebox{1}{
\begin{tabular}{lclc}
\toprule
Methods   & $P_{acc}$   & Methods   & $P_{acc}$    \\ 
\midrule
\multicolumn{4}{c}{\textbf{\textit{Traditional machine learning methods}}} \\ 
\midrule
KNN\cite{zhou2023pr} & 60.66/07.93
&CORAL\cite{zhou2023pr}& 68.15/07.83\\
Adaboost\cite{zhou2023pr}& \underline{71.87/05.70}
&RF\cite{MetaEmotionNet2024} &53.30/08.70\\
SVM\cite{MetaEmotionNet2024} &53.10/10.90
&TCA\cite{CAI2024} &63.34/14.88\\
SA\cite{CAI2024} &69.00/10.89
&GFK\cite{CAI2024} &71.31/14.09\\
\midrule
\multicolumn{4}{c}{\textit{\textbf{Deep learning methods}}}\\ 
\midrule
TSception\cite{MetaEmotionNet2024}&64.30/09.80
&CDCN\cite{MetaEmotionNet2024} &69.30/09.40\\
MLP\cite{MetaEmotionNet2024} &70.10/09.40
&STFFNN\cite{MetaEmotionNet2024} &72.00/09.00\\
STRNN\cite{MetaEmotionNet2024}&73.20/11.10
&3D-CNN\cite{MetaEmotionNet2024}&74.20/07.80\\
MMResLSTM\cite{MetaEmotionNet2024} &74.40/10.40
&ACRNN\cite{MetaEmotionNet2024} &76.30/08.10 \\
MetaEmotionNet\cite{MetaEmotionNet2024} &77.50/08.80
&A-LSTM\cite{CAI2024} &72.18/10.85 \\
DGCNN\cite{CAI2024} &79.95/09.02
&EEG-SWTNS\cite{CAI2024} &80.07/10.75\\
\midrule
\multicolumn{3}{l}{\textbf{M3D}} & \textbf{77.05/07.61}  \\
\bottomrule
\end{tabular}
}
\end{center}
\end{table}

\begin{table}[]
\begin{center}
\caption{Performance comparison on SEED-IV using cross-subject cross-session LOSO CV. The \underline{second-best} traditional machine learning method result is underlined.}
\label{tab:seedIV_crossCompare}
\setlength{\tabcolsep}{1.8mm}
\scalebox{1}{
\begin{tabular}{lclc}
\toprule
Methods   & $P_{acc}$   & Methods   & $P_{acc}$    \\ 
\midrule
\multicolumn{4}{c}{\textbf{\textit{Traditional machine learning methods}}} \\ 
\midrule
KPCA\cite{li2018cross} &51.76/12.89
&TPT\cite{li2018cross} &52.43/14.43\\
KNN\cite{zhou2023pr} & 40.83/07.28
&Adaboost\cite{zhou2023pr} & 53.44/09.12\\
CORAL\cite{zhou2023pr} &49.44/09.09
&RF\cite{MetaEmotionNet2024} &34.70/06.60\\
SVM\cite{MetaEmotionNet2024} &41.10/07.40
&TCA\cite{CAI2024} &56.56/13.77\\
SA\cite{CAI2024} &\underline{64.44/09.46}
&GFK\cite{CAI2024} &64.38/11.41\\
\midrule
\multicolumn{4}{c}{\textit{\textbf{Deep learning methods}}}\\ 
\midrule 
DANN\cite{li2018cross}  & 54.63/08.03
&DAN \cite{li2018cross}   & 58.87/08.13\\
MLP\cite{MetaEmotionNet2024} &50.70/06.30
&STFFNN\cite{MetaEmotionNet2024} &56.70/07.30 \\
ACRNN\cite{MetaEmotionNet2024} &49.20/09.20 
&MMResLSTM\cite{MetaEmotionNet2024} &51.10/11.40\\
TSception\cite{MetaEmotionNet2024}&56.20/09.50
&STRNN\cite{MetaEmotionNet2024} &53.20/07.40\\
3D-CNN\cite{MetaEmotionNet2024} &54.10/10.90
&CDCN\cite{MetaEmotionNet2024} &54.50/13.10\\
MetaEmotionNet\cite{MetaEmotionNet2024} &61.20/08.30
&A-LSTM\cite{CAI2024}&55.03/09.28\\
DGCNN\cite{CAI2024}& 52.82/09.23 
&EEG-SWTNS\cite{CAI2024}&66.72/10.19\\
\midrule
\multicolumn{3}{l}{\textbf{M3D}} & \textbf{64.90/10.23}  \\
\bottomrule
\end{tabular}
}
\end{center}
\end{table}

\begin{table}[]
\begin{center}
\caption{Performance comparison on SEED-V using cross-subject cross-session LOSO CV. The \underline{second-best} traditional machine learning method result is underlined. Models reproduced by the authors are indicated with an asterisk (*).}
\label{tab:seedV_crossCompare}
\setlength{\tabcolsep}{2.8mm}
\scalebox{1}{
\begin{tabular}{lclc}
\toprule
Methods   & $P_{acc}$   & Methods   & $P_{acc}$    \\ 
\midrule
\multicolumn{4}{c}{\textbf{\textit{Traditional machine learning methods}}} \\ 
\midrule
TCA*\cite{5640675} &34.15/08.04
&SA*\cite{SA2013} &41.20/11.95 \\
GFK*\cite{gong2012geodesic} &35.94/09.51
&SVM*\cite{1995Support} &46.72/12.50 \\
Adaboost*\cite{1997A} &\underline{47.46/12.82}
&KNN*\cite{coomans1982alternative} &35.00/09.38\\
CORAL*\cite{2015Return} &46.23/12.58
&KPCA*\cite{mika1998kernel} &32.23/12.09 \\
\midrule
\multicolumn{4}{c}{\textit{\textbf{Deep learning methods}}}\\ 
\midrule
DGCNN\cite{li2024mslte}&37.38/08.64 &DANN \cite{li2024mslte}&46.87/10.09\\
ADDA\cite{li2024mslte} &47.97/09.43 &GMSS\cite{li2024mslte} &46.34/09.65\\
MSLTE\cite{li2024mslte} &50.12/08.59 
&A-LSTM\cite{zhou2023progressive} &40.34/08.68\\
BDGLS\cite{zhou2023progressive} &59.59/04.75 & IAG\cite{zhou2023progressive} &59.68/09.44\\
ECLGCNN\cite{zhou2023progressive} &61.58/10.35
&DAN*\cite{li2018cross} &59.32/10.30\\
DCORAL*\cite{sun2016deep}  &57.41/09.65
&DDC*\cite{2014Deep} &59.44/07.04\\
\midrule
\multicolumn{3}{l}{\textbf{M3D}} & \textbf{59.15/09.30}  \\
\bottomrule
\end{tabular}
}
\end{center}
\end{table}

\subsection{Experimental Results on Cross-Subject Single-Session}
The experimental results on SEED, SEED-IV and SEED-V database under the cross-subject single-session cross-validation are reported in Table \ref{tab:seed_and_seediv and_seedv_singlesession}. For comprehensive evaluation, we compare our method with existing approaches under identical experimental conditions. The comparison methods included in each table are selected based on the availability of reported results under the corresponding dataset and evaluation protocol to ensure fair and consistent evaluation. The baseline methods selected for comparison include a range of classical and recent approaches, covering representative state-of-the-art techniques published in the field of EEG-based domain adaptation. Results marked with * are independently reproduced by the authors under standardized preprocessing and evaluation protocols to ensure comparability.

For the SEED database, the best performance is obtained when DT is adopted as the initial classifier, where the corresponding accuracy, specificity and precision are 84.57\%, 88.78\% and 76.94\%. We compare the obtained experimental results with the existing literature, as shown in Table \ref{tab:seed_singleCompare}. The comparison results show that the proposed non-deep transfer learning model (M3D) demonstrates superior performance. In comparison to the best results obtained by the traditional machine learning methods, our model shows a performance enhancement of 8.26\%. Furthermore, it performs comparably or even surpasses some deep learning methods, such as  ADA \cite{2019Domain}, TANN \cite{li2021novel}, P-GCNN \cite{pan2023st} and DG-DANN \cite{ma2019reducing}.

For the SEED-IV database, the best accuracy is obtained when DT is used as the initial classifier, where the corresponding accuracy, sensitivity, F1-score, AUROC and NPV are 60.94\%, 78.10\%, 50.49\%, 73.68\% and 88.02\%, respectively. The performance comparison results with the existing methods are reported in Table \ref{tab:seedIV_singleCompare}. Compared to the best results reported in the traditional machine learning methods, the proposed M3D demonstrates a slight performance advantage of 0.72\%. Furthermore, it performs comparably or even surpasses some deep learning methods, including ScalingNet \cite{cheng2024emotion}and MS-MDA \cite{2021MS}.

For the SEED-V database, the best accuracy is obtained when DT is used as the initial classifier, where the corresponding accuracy, specificity, precision, F1-score and AUROC are 65.25\%, 66.16\%, 16.6\%, 25.76\% and 78.46\%. As shown in Table \ref{tab:seedV_singleCompare}, the M3D model outperforms the best-performing traditional machine learning baseline by a notable 14.15\%, clearly demonstrating its superior classification capability.

\subsection{Experimental Results on Cross-Subject Cross-Session}
To evaluate the efficiency and stability of the proposed M3D in managing variations in EEG data collected from the same subject at different times, we also conduct a model validation under the cross-subject cross-session leave-one-subject-out cross-validation protocol. The experimental results are reported in Table \ref{tab:seed_and_seediv_and_seedv_crosssession}, when different initial classifiers are used. The best model performance of SEED database is achieved when DT is adopted as the initial classifier, where the accuracy is 77.05\%, and the other six indicators, except for NPV, are also the highest. The performance results against the existing methods are compared in Table \ref{tab:seed_crossCompare}. It shows, the proposed model demonstrates an accuracy improvement of 5.18\%  in comparison to traditional machine learning methods. Further, it exhibits performance that is comparable, to some extent, with deep learning methods, showing the effectiveness of the proposed M3D in managing variations in data collected from the same subject across different time periods.

\begin{table}[]
\begin{center}
\caption{Cross-subject single-session 10 fold cross-validation on MDD with different initial classifiers.}
\label{tab:mdd_10fold_singlesession}
\setlength{\tabcolsep}{1.2mm}
\scalebox{1}
{
\begin{tabular}{llcccccc}
\toprule
Session& &KNN & SVM  & DT & Adaboost & GNB & Bagging   \\ 
\midrule
\multirow{7}{*}{EC} &\multicolumn{1}{|l}{Accuracy} & \textbf{82.72}  &\textbf{82.72}&\textbf{82.72}&\textbf{82.72} & 81.06&81.06  \\
&\multicolumn{1}{|l}{Sensitivity} &\textbf{81.86} &\textbf{81.86}&\textbf{81.86}&\textbf{81.86} &78.99 &78.99\\
&\multicolumn{1}{|l}{Specificity} &\textbf{82.45}&\textbf{82.45}&\textbf{82.45}&\textbf{82.45}&81.81 &81,81\\
&\multicolumn{1}{|l}{Precision} &\textbf{81.93}&\textbf{81.93}&\textbf{81.93}&\textbf{81.93}&\textbf{81.93}&\textbf{81.93}\\
&\multicolumn{1}{|l}{F1-score}  & \textbf{81.89} &\textbf{81.89}  & \textbf{81.89}  &\textbf{81.89}  &80.43&80.43  \\
&\multicolumn{1}{|l}{AUROC} &\textbf{82.15}&\textbf{82.15}&\textbf{82.15}&\textbf{82.15} &80.40 &80.40\\
&\multicolumn{1}{|l}{NPV} &\textbf{82.38} &\textbf{82.38}&\textbf{82.38}&\textbf{82.38} &78.86 &78.86\\
\midrule
\midrule
\multirow{7}{*}{EO} &\multicolumn{1}{|l}{Accuracy} & 74.96    & 74.96 & \textbf{78.63} & 74.96&74.96 & 74.96 \\
&\multicolumn{1}{|l}{Sensitivity} &73.98&73.98&\textbf{77.68}&73.98&73.98&73.98\\
&\multicolumn{1}{|l}{Specificity} &75.79&75.79 &\textbf{79.24} &75.79&75.79&75.79\\
&\multicolumn{1}{|l}{Precision} &73.97&73.97 &\textbf{77.67}&73.97&73.97&73.97\\
&\multicolumn{1}{|l}{F1-score}  & 73.97 &73.97& \textbf{77.67} & 73.97 &73.97 &73.97 \\
&\multicolumn{1}{|l}{AUROC} &74.89&74.89 &\textbf{78.46} &74.89&74.89&74.89\\
&\multicolumn{1}{|l}{NPV} &75.80 &75.80 &\textbf{79.25} &75.80 &75.80 &75.80\\
\bottomrule
\end{tabular}
}
\end{center}
\end{table}

\begin{table}
\caption{Performance comparison on MDD using cross-subject 10-fold cross-validation. EC: closed-eye; EO: open-eye. The \underline{second-best} traditional machine learning method result is underlined. Models reproduced by the authors are indicated with an asterisk (*).}
\label{tab:mdd_10foldCompare}
    \centering
    \renewcommand{\arraystretch}{1}
    \resizebox{\columnwidth}{!}{%
    \setlength{\tabcolsep}{7mm}{
        \begin{tabular}{lcc}
        \toprule
        \multirow{2}{*}{Methods}  & EC session  & EO session \\ &$P_{acc}$&$P_{acc}$  \\ 
        \midrule
        \multicolumn{3}{c}{\textit{\textbf{Traditional machine learning methods}}}\\ 
        \midrule
        SVM*\cite{1995Support} &35.56/10.59  & 35.56/10.59 \\
        CORAL*\cite{2015Return}    & 60.27/15.25 & 60.12/13.58\\
        RF*\cite{Breiman2001Random}       &69.44/19.57  &69.44/19.57  \\
        Adaboost*\cite{1997A} & 70.49/16.70 & 71.00/19.62\\
        KNN*\cite{coomans1982alternative}   & 72.25/18.32 &  72.25/18.32\\
        SA*\cite{SA2013}       &\underline{74.21/11.98}   & \underline{74.21/11.98}\\
        \midrule
        \multicolumn{3}{c}{\textit{\textbf{Deep learning methods}}}\\ 
        \midrule
        DAN* \cite{li2018cross}  & 64.19/05.30  & 64.99/09.01 \\
        DDC*\cite{2014Deep}   & 64.65/04.84  & 65.69/08.08 \\
        DANN*\cite{li2018cross}  & 66.49/11.08  & 70.25/09.67 \\
        DCORAL*\cite{sun2016deep} & 70.88/10.53  & 74.49/08.17 \\
        \midrule
        \textbf{M3D} & \textbf{82.72/11.35} & \textbf{78.63/13.79}  \\
        \bottomrule
        \end{tabular}}}
\end{table}

For the SEED-IV database, the performance across various initial classifiers is relatively consistent, with the highest accuracy reaching 64.9\% when DT is employed as the initial classifier. Table \ref{tab:seedIV_crossCompare} reports the performance comparison results against the existing machine learning and deep learning methods. Compared to the best results reported in various literature on machine learning methods, the proposed M3D shows a performance improvement of 0.46\%. Furthermore, it performs comparably or even outperforms some deep learning methods, including MetaEmotionNet \cite{MetaEmotionNet2024} and DAN \cite{li2018cross}. The superior cross-session validation performance observed in SEED-IV is largely driven by its smaller per-session sample size, higher emotional class complexity, and more diverse multimodal stimuli, which collectively benefit from cross-session aggregation. Meanwhile, the SEED database’s more stable per-session paradigm and stimulus uniformity make single-session validation more effective and reliable.

\begin{table*}[!h]
\centering
\caption{Performance comparison without classifier learning and ensemble learning, with only classifier learning, and with both classifier learning and ensemble learning.}
\label{tab:seed_singlesession_abaltion_study}
\begin{adjustbox}{width=\textwidth}
\large 
\renewcommand\arraystretch{1.2}
\begin{tabular}{@{}cccccccccc@{}}
\toprule
Manifold Feature Transformation & \makecell{Dynamic Distribution Alignment \\ and Classifier Learning} &Ensemble Learning & Accuracy &Sensitivity &Specificity &Precision & F1-score &AUROC &NPV\\
\midrule
\checkmark &  &  & 48.27/14.39 &37.70 &54.65 &33.40&35.42 &61.52 &59.26 \\
 & \checkmark & &69.51/16.40&62.64 &72.69 &51.51 & 56.53 &77.16 &80.77\\
  & \checkmark &\checkmark &71.01/13.42 &65.17 &73.83 &54.58& 59.41 &78.25 &81.46\\
 \checkmark & \checkmark & &79.66/09.34 &74.37 &82.21 &66.86 &70.41 &84.68&86.92\\
 \midrule
 \checkmark & \checkmark &\checkmark &84.57/09.49&76.04 &88.78 &76.94 &76.49 &88.36 & 88.27 \\
\bottomrule
\end{tabular}
\end{adjustbox}
\end{table*}


For the SEED-V database in the cross-session setting, performance across various initial classifiers remains relatively stable, with the highest accuracy reaching 59.15\%, again observed with DT. As shown in Table \ref{tab:seedV_crossCompare}, the proposed M3D model outperforms the best results reported in the literature for traditional machine learning methods by a margin of 11.69\%, demonstrating a substantial improvement in generalization capability. Furthermore, M3D delivers performance that is comparable to or even surpasses certain deep learning models, including the state-of-the-art DCORAL method \cite{sun2016deep}.


\subsection{Experimental Results on clinical EEG database}
To assess the generalizability of the proposed M3D model in real-world clinical applications, we evaluate its performance on the MDD database (n=56). Following previous work \cite{ye2024semi}, we adopt the cross-subject single-session 10-fold cross-validation protocol for fair comparison. The validation results are presented in Table \ref{tab:mdd_10fold_singlesession}, demonstrating consistent performance across various initial classifiers. In particular, the highest accuracy is achieved using DT as the initial classifier, reaching 82.72\% in the EC session and 78.63\% in the EO session. For comparative analysis, we report performance differences against existing methods in Table \ref{tab:mdd_10foldCompare}. The proposed M3D model outperforms the best-performing traditional machine learning approach by 8.51\% and 4.42\% in the EC and EO sessions, respectively. Among deep learning methods, DCORAL \cite{sun2016deep} achieves the highest reported accuracy in prior studies, with 70.88±10.53\% in the EC session and 74.49±8.17\% in the EO session. M3D surpasses these results by 11.84\% and 4.14\% in the EC and EO sessions, respectively, demonstrating its superior ability to capture the complex patterns associated with MDD. These results underscore the effectiveness of M3D in real-world clinical settings, establishing it as a promising tool for practical clinical applications.

\section{Discussion}
\subsection{Ablation Study}
To comprehensively examine the contributions of the model components within the proposed M3D model, we conduct a thoroughly ablation study. Specifically, we design four model variations to isolate the effects of key modules. \textbf{(1) Manifold Feature Transformation Only:} the initial decision tree classifier is trained to evaluate model performance after applying only the manifold feature transformation module. \textbf{(2) Dynamic Distribution Alignment and Classifier Learning Only:} the DE feature is used as input, focusing solely on dynamic distribution alignment and classifier learning. \textbf{(3) Without Manifold Feature Transformation:} the DE feature serves as input for dynamic distribution alignment, classifier learning, and ensemble learning, excluding the manifold feature transformation module. \textbf{(4) Without Ensemble Learning:} the ensemble learning module is removed to evaluate its impact on model performance. Note here that all the presented experimental results in this section are based on the SEED database using cross-subject single-session leave-one-out cross-validation. The decision tree is employed as the initial classifier, with 10 iterative loops for evaluation.

Table \ref{tab:seed_singlesession_abaltion_study} presents the results of the ablation study, demonstrating that the proposed M3D model achieves the best performance by effectively integrating all modules. Using only the manifold feature transformation module leads to the lowest performance, with a classification accuracy of 44.73±14.39\%.  Focusing on dynamic distribution alignment and classifier learning improves the accuracy to 69.51±16.4\%. Adding ensemble learning to dynamic distribution alignment and classifier learning further increases the accuracy to 71.01±13.42\%, a 1.5\% improvement, highlighting the effectiveness of ensemble learning in EEG-based emotion recognition. However, removing the manifold feature transformation module causes a significant accuracy drop from 84.57±9.49\% (full model) to 71.01±13.42\%, emphasizing the importance of feature transformation before distribution alignment. Removing the ensemble learning module also reduces the accuracy to 79.66±9.34\%. Additionally, features in the Grassmann manifold space prove more robust than those in the original feature space, improving distribution alignment. All four ablation models significantly differ from the full model (p $<$ 0.0001). We use an independent t-test for normally distributed groups and the Wilcoxon Rank Sum test for nonnormally distributed ones, with FDR correction applied to enhance result reliability.

\begin{figure}[]
\begin{center}
\subfloat[]{\includegraphics[width=0.5\textwidth]{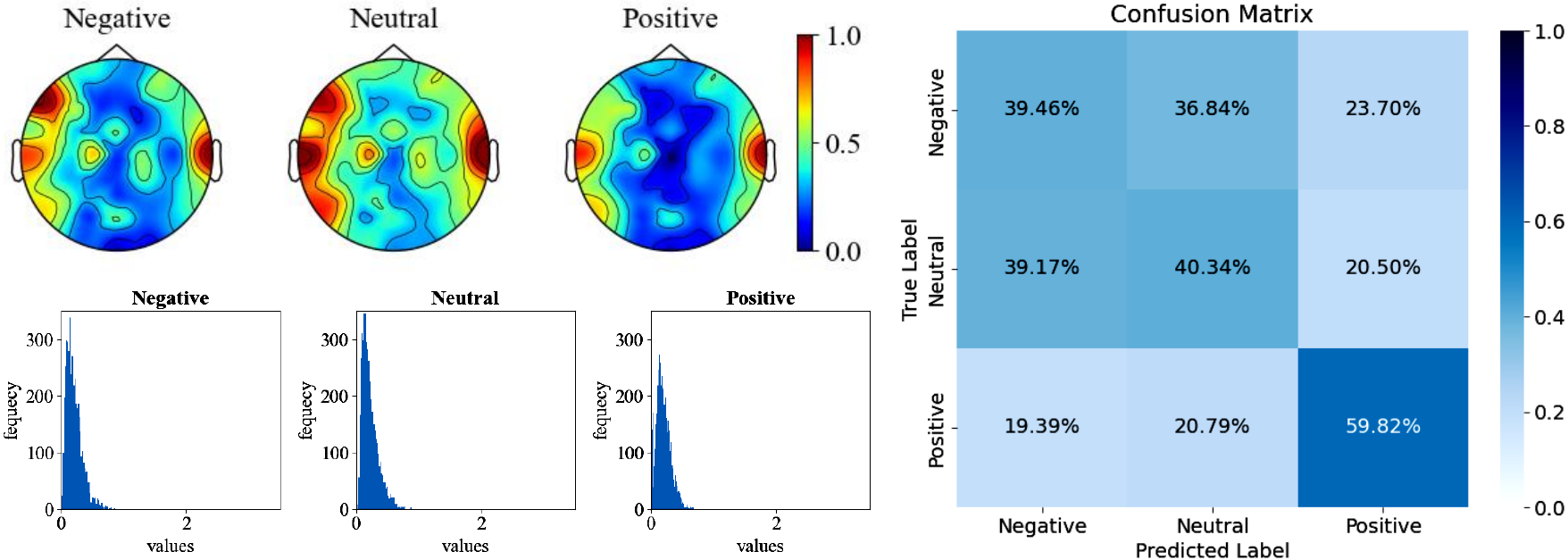}}

\subfloat[]
{\includegraphics[width=0.5\textwidth]{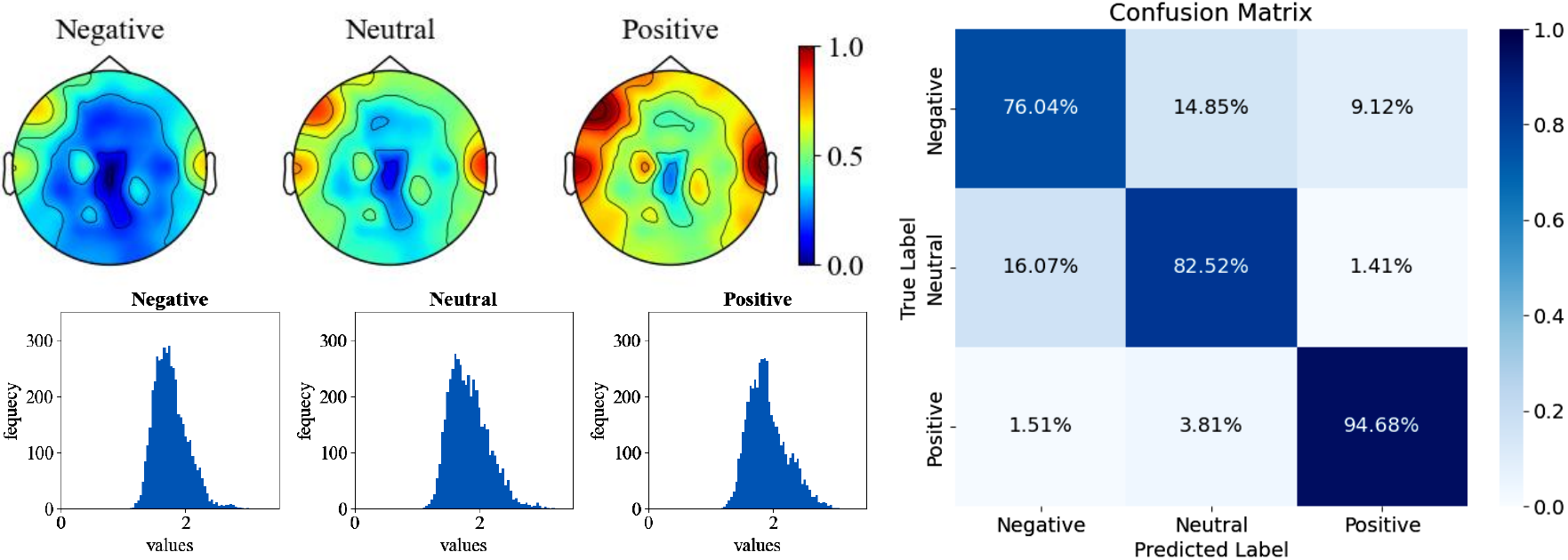}}
\caption{Visualization of topographic analysis, histogram distribution of mutual information between EEG patterns and model predictions, and confusion matrices before and after alignment (a) before alignment, (b) after alignment.}
\label{fig:topo}
\end{center}
\end{figure}

\subsection{The effect of the parameter}
We conduct a series of experiments under the cross-subject single-session setting using the SEED database to investigate the impacts of key parameters (subspace dimension $d$, regularization parameters $\eta$, $\rho$, $\lambda$), the number of iterations $\iota$, and the training sample size on the performance of the M3D model. The results show that the model achieves the best performance when the subspace dimension $d=128$, and it demonstrates good robustness to variations in the regularization parameters within specific ranges. The model's classification performance fluctuates at the beginning of iterations and stabilizes after $\iota=10$, demonstrating the training advantage of M3D in cross-domain tasks. Using only 3,000 training samples, the model achieves an accuracy of 83.11\% with a runtime of 37.4 seconds, outperforming several deep learning methods that use all available training samples. For instance, GECNN \cite{pan2023st} achieves an accuracy of 82.46±10.83\% and MS-MDA \cite{she2023multisource} obtains 79.67±8.01\%. More detailed descriptions and results are presented in Supplementary Materials Appendix C. And we provide a comprehensive theoretical analysis of the computational complexity of the proposed M3D framework in Supplementary Materials Appendix D.

\subsection{Investigating factors affecting model performance}
We also conduct a series of experiments to study the factors that affect the performance of the model. First, replacing the Grassmann manifold with the Stiefel manifold in the manifold feature transformation module leads to a decrease in the model’s accuracy to 67.94\%, demonstrating the Grassmann manifold's superiority in extracting EEG emotion features. Second, comparing TCA with Principal Component Analysis (PCA) in manifold feature transformation, TCA shows better performance in terms of accuracy and F1-score, as it can map, reduce dimensions, and maximize the similarity between domains simultaneously. Third, experiments with fixed $\mu$ values in dynamic distribution alignment show that the adaptive $\mu$ significantly improves the model performance, increasing the accuracy by 4.26\%. This mechanism mimics the brain's ability to prioritize different information streams based on contextual discrepancies, representing a more biologically plausible and data-responsive adaptation process. Finally, the evaluation of ensemble learning indicates that both the voting method and the LinkCluE method can enhance the reliability of classification results, with the LinkCluE method outperforming the voting method. More detailed descriptions and results are presented in the Supplementary Materials Appendix E.

\subsection{Visualization of Feature Alignment}
We employ t-SNE visualizations and EEG topographic maps to intuitively demonstrate the alignment of source and target features before and after adaptation. For t-SNE \cite{van2008visualizing} visualization, we project the source and target features into a two-dimensional space before and after alignment. Distinct colors represent different emotion categories, while different shapes indicate domain origin (source vs. target). The visualization (Supplementary Materials Appendix F, Fig. S13) clearly demonstrates that after alignment, the distributional discrepancy between the source and target domains is significantly reduced, while the separability between different emotion classes improves.

Additionally, we conduct topographic analysis by computing the mutual information between EEG patterns and prediction labels to identify important brain patterns for emotion recognition. The mutual information between the input features of the target domain and the model prediction results is estimated using a non-parametric method \cite{1987sample,kraskov2004estimating,ross2014mutual}. After normalizing the mutual information matrix, we visualize the average mutual information before and after alignment. The results in Fig. \ref{fig:topo} show that the differences among the EEG topographic maps of the three emotions increase after alignment. The confusion matrix further supports this finding, demonstrating that the model achieves better classification performance following alignment. Moreover, the analysis reveals that the EEG patterns with higher informativeness for emotion recognition are mainly located in the prefrontal region \cite{zhong2020eeg, 2019Identifying}. More detailed descriptions and results are presented in the Supplementary Materials Appendix F.

\section{Conclusion}
In this paper, we propose a novel non-deep transfer learning framework, termed M3D, for cross-subject and cross-session EEG-based emotion recognition, as well as clinical mental disorder recognition. The framework employs a manifold-based domain adaptation with dynamic distribution alignment, adaptively balancing marginal and conditional distributions in the Grassmann manifold space. This enhances intrinsic data representation and reduces domain disparity. Additionally, optimized classifier learning with ensemble techniques improves robustness and reliability in handling EEG data variations. Extensive experiments on three benchmark databases validate the framework under two evaluation protocols. Furthermore, we demonstrate the model’s effectiveness on a real clinical MDD dataset. Compared to existing methods, M3D achieves promising results against non-deep approaches and performance comparable to deep learning models, with an average improvement of 6.67\% in classification accuracy. These findings highlight the potential of non-deep transfer learning in addressing individual and session variations in affective brain-computer interfaces. On the other hand, in the dynamic distribution alignment module, we did not impose constraints on the adaptive factor $\mu$ during the iteration process, which may lead to suboptimal feature alignment in certain cases. Addressing this limitation in future work could further enhance the stability and generalizability of the framework. And we plan to investigate the adaptability of the M3D framework to more diverse clinical EEG datasets (e.g., epilepsy monitoring or sleep staging) and to evaluate the stability of model performance under varying preprocessing pipelines.


\appendices
\section{Proof of the computation $\mu$}
The a daptive factor $\mu$ is designed to dynamically adjust the emphasis between marginal and conditional distribution adaptation based on the differences between the source and target domains. We use the A-distance measurement \cite{ben2006analysis} as a metric to estimate these differences, which is a well-established measure in the domain adaptation literature. The adaptive factor $\mu$ is computed as
\begin{equation}
    \mu=1-\dfrac{d_{A}}{d_{A}+\sum^C_{c=1}d_{c}},
\end{equation}
where $d_{A}$ is the marginal distribution difference between the source and target domains and $d_{c}$ is the conditional distribution difference between the class $c$ ($c \in [1,\dots, C]$) from the source and target domains. When $d_{A}$ is large compared to $\sum^C_{c=1}d_{c}$ , it means the marginal distribution difference between the source and target domains is significant. In this case, $\mu$ will be close to 0, indicating that more weight should be placed on marginal distribution adaptation. Conversely, when $\sum^C_{c=1}d_{c}$  is large relative to $d_{A}$, $\mu$ will be close to 1, suggesting that conditional distribution adaptation is more crucial.

For the marginal distribution difference, the A-distance $d_{A}$ provides a quantification of the marginal distribution difference between the source domain $T_{S}$ and the target domain $T_{T}$, and is defined as
\begin{equation}
    d_{A}\left(T_{S},T_{T}\right)=2\left(1-2\epsilon\left(h\right)\right),
\end{equation}
where $\epsilon\left(h\right)$ denotes the hinge loss, which is derived from a binary classifier trained specifically to distinguish between samples from the source and target domains. A smaller A-distance indicates a closer marginal distribution.

For the conditional distribution difference, based on the provided label information for source data and the estimated pseudo-label information of target data, we conduct the A-distance measurement within each $c$ class as
\begin{equation}
    d_{c}=d_{A}(T^{{c}}_S,T^{(c)}_T),
\end{equation}
Here, $T_{S}^{(c)}$ and $T_{T}^{(c)}$ represent the subsets of data corresponding to class $c$ from the source and target domains, respectively. By aggregating these class-specific A-distances ($\sum^C_{c=1}d_{c}$) and comparing them with the marginal A-distance $d_{A}$, we can determine the relative importance of marginal and conditional distribution adaptation.

\section{Statistically analyze EEG signal differences across subjects}
To thoroughly investigate EEG signal differences across subjects in the SEED and SEED-IV datasets, we perform a comprehensive statistical analysis on the input EEG features, which includes both qualitative visualizations and quantitative comparisons. Specifically, our analysis includes three main parts: (1) low-dimensional feature visualization using t-Distributed Stochastic Neighbor Embedding (t-SNE) to reveal the overall data distribution and inter-subject separability; (2) box plot analysis to display the statistical distribution of EEG features for each subject across different emotional states; and (3) statistical difference matrices constructed through rigorous hypothesis testing to quantify the inter-subject variability under the same emotion condition.

\begin{figure*}[h]
\begin{center}
\includegraphics[width=1\textwidth]{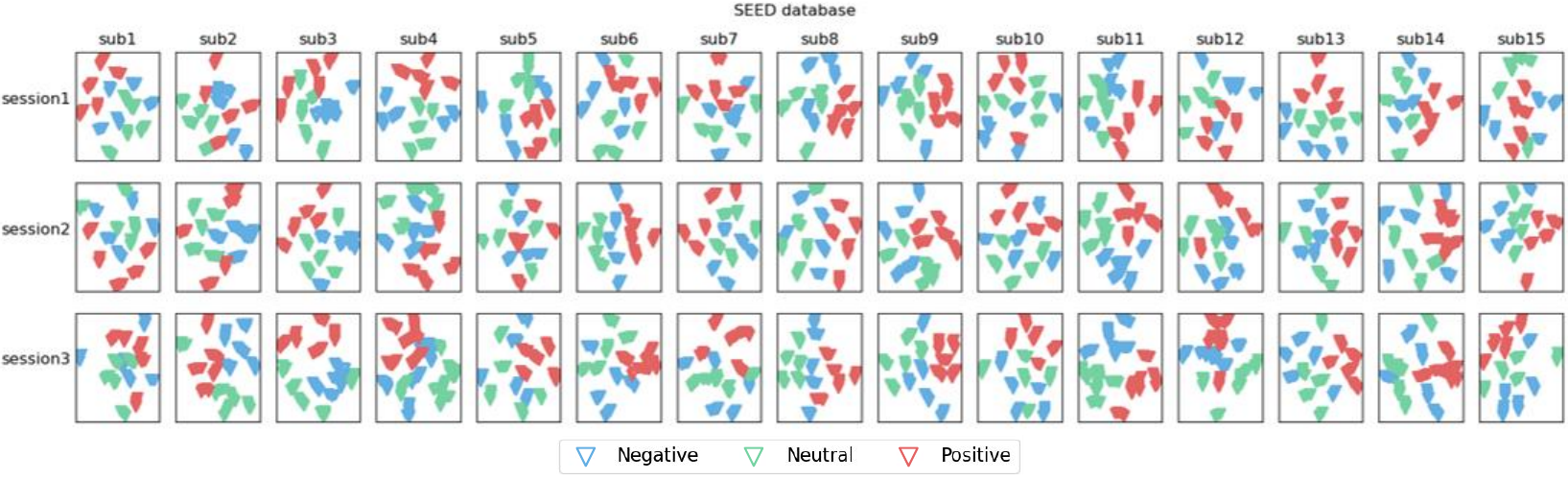}
\caption{A t-SNE visualization of the input EEG features in SEED database. The blue, green, and red colors indicate negative, neutral, and positive emotions.}
\label{fig:seed_rawdata}
\end{center}
\end{figure*}

\begin{figure*}[h]
\begin{center}
\includegraphics[width=1\textwidth]{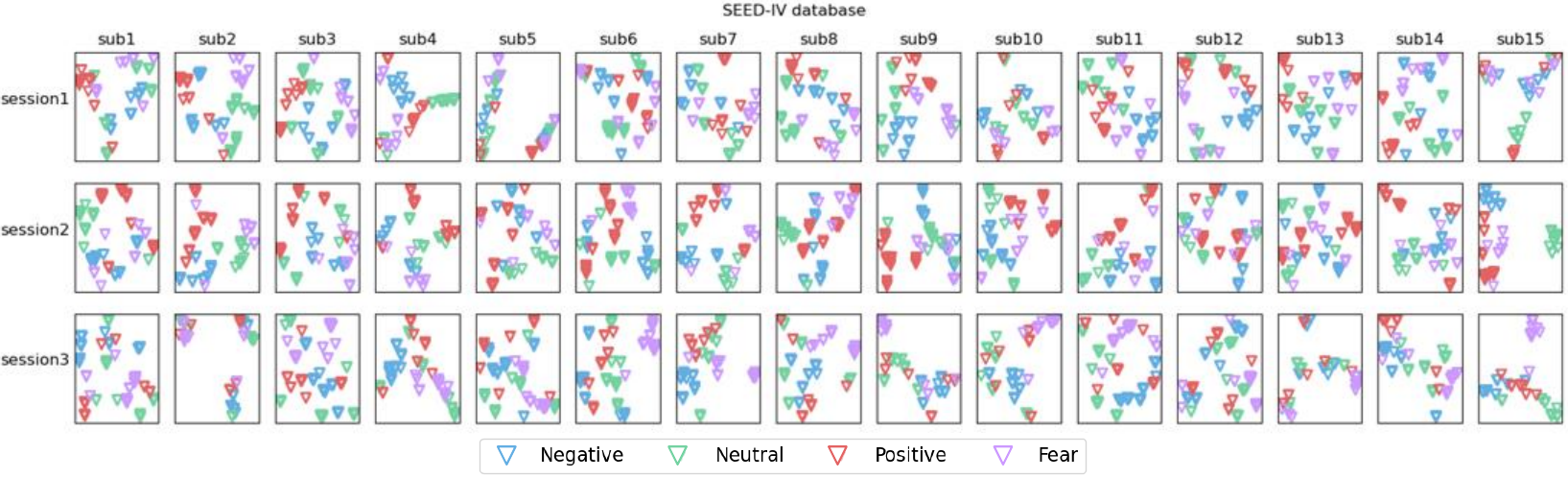}
\caption{A t-SNE visualization of the input EEG in SEED-IV database. The blue, green, red and purple colors indicate negative, neutral, positive and fear emotions.}
\label{fig:seediv_rawdata}
\end{center}
\end{figure*}

Firstly, we apply t-SNE to visualize the distribution of the input EEG features in a two-dimensional space. The results (Fig. \ref{fig:seed_rawdata} and \ref{fig:seediv_rawdata}) clearly show distinct clustering patterns for different emotional states and reflect the variability of feature distributions across subjects. This visualization offers an intuitive understanding of how subject-specific factors impact feature separability and emotional clustering.

\begin{figure}[h]
\begin{center}
\includegraphics[width=1\linewidth]{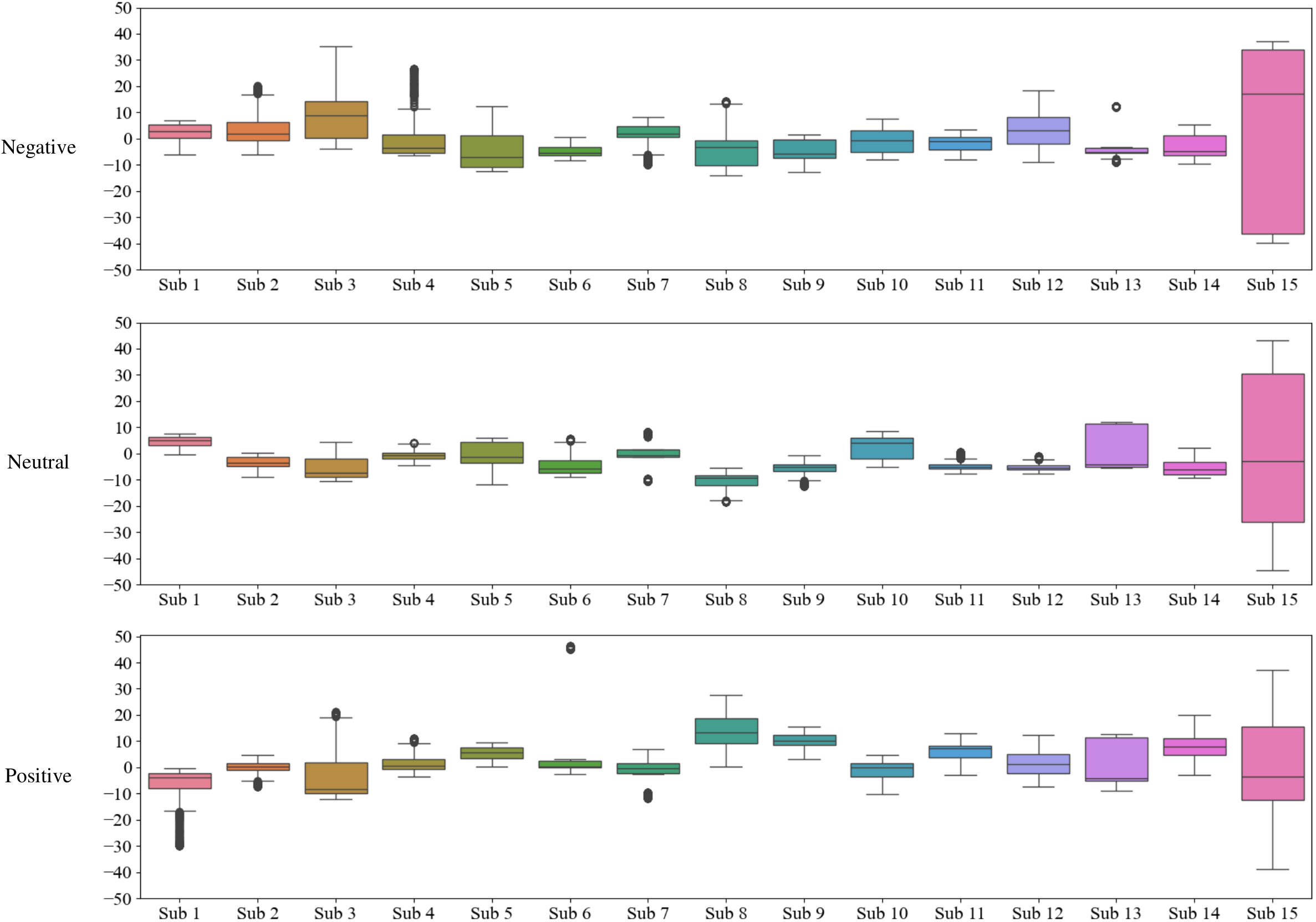}
\caption{Box plots of the input EEG feature distribution across 15 subjects and 3 emotions (negative, neutral, positive) in a single session of SEED database.}
\label{fig:SEED_statistical}
\end{center}
\end{figure}

\begin{figure}[h]
\begin{center}
\includegraphics[width=1\linewidth]{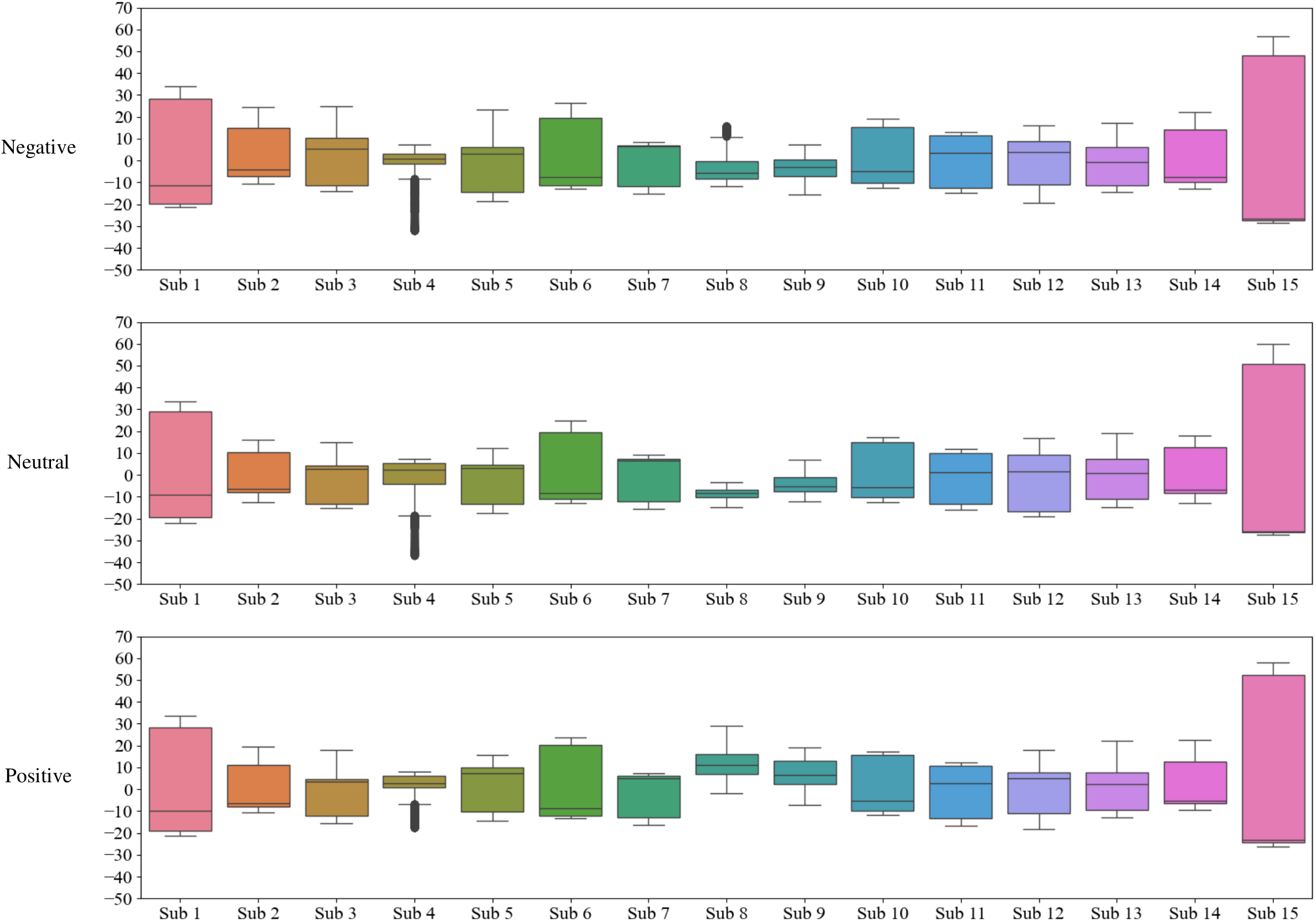}
\caption{Box plots of the input EEG feature distribution across 15 subjects and 3 emotions (negative, neutral, positive) in the cross session of SEED database.}
\label{fig:SEED_statistical_3session}
\end{center}
\end{figure}

\begin{figure}[h]
\begin{center}
\includegraphics[width=1\linewidth]{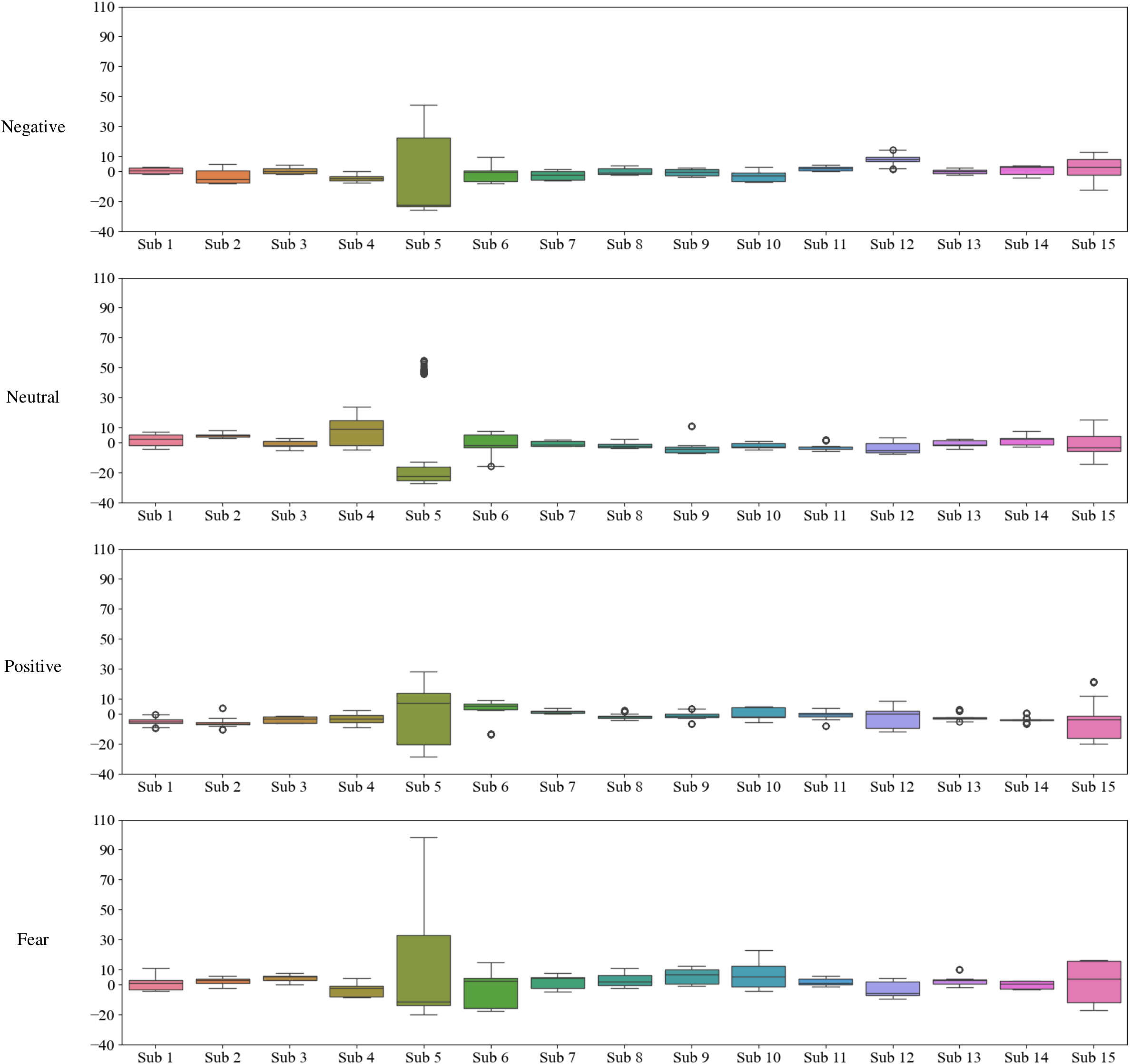}
\caption{Box plots of the input EEG feature distribution across 15 individuals and 4 emotions (negative, neutral, positive, fear) in a single session of SEED-IV database.}
\label{fig:SEEDIV_statistical}
\end{center}
\end{figure}

\begin{figure}[h]
\begin{center}
\includegraphics[width=1\linewidth]{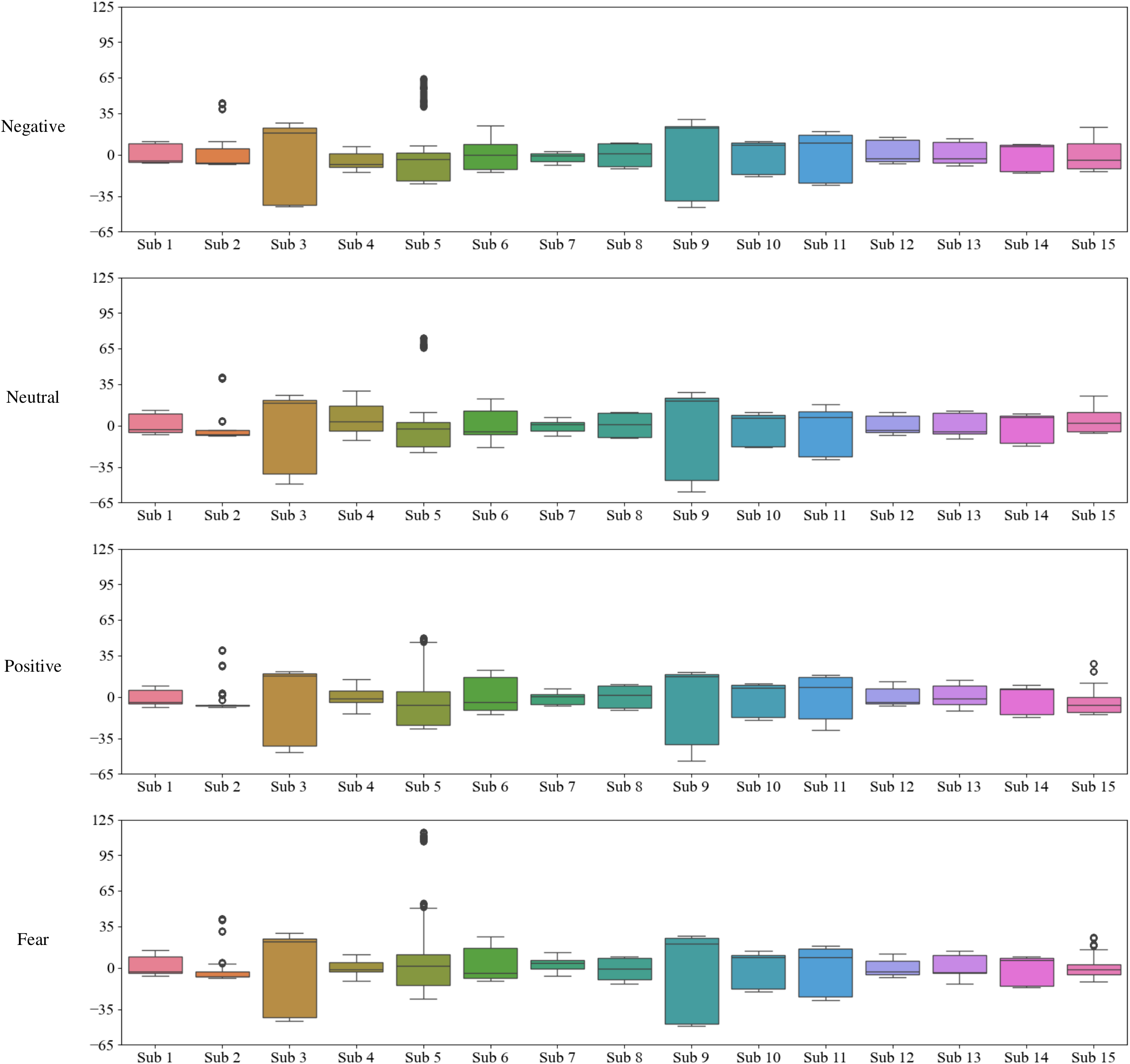}
\caption{Box plots of the input EEG feature distribution across 15 individuals and 4 emotions (negative, neutral, positive, fear) in the cross session of SEED-IV database.}
\label{fig:SEEDIV_statistical_3session}
\end{center}
\end{figure}

Secondly, we perform feature dimensionality reduction followed by box plot analysis for each subject and emotion category (Fig. \ref{fig:SEED_statistical}-\ref{fig:SEEDIV_statistical_3session}). The box plots provide detailed statistical information, including medians, interquartile ranges, and outliers, and the analysis results reflect the extent of variation in input EEG features among individuals. We observe that, even within the same emotional condition, the input EEG feature distributions vary significantly across subjects, suggesting that individual differences play a substantial role in neural responses to emotional stimuli, and further, an enlarged intra-subject variance is observed under cross-session settings, indicating temporal fluctuations in individual neural patterns.

\begin{figure*}[h]
\begin{center}
\includegraphics[width=1\textwidth]{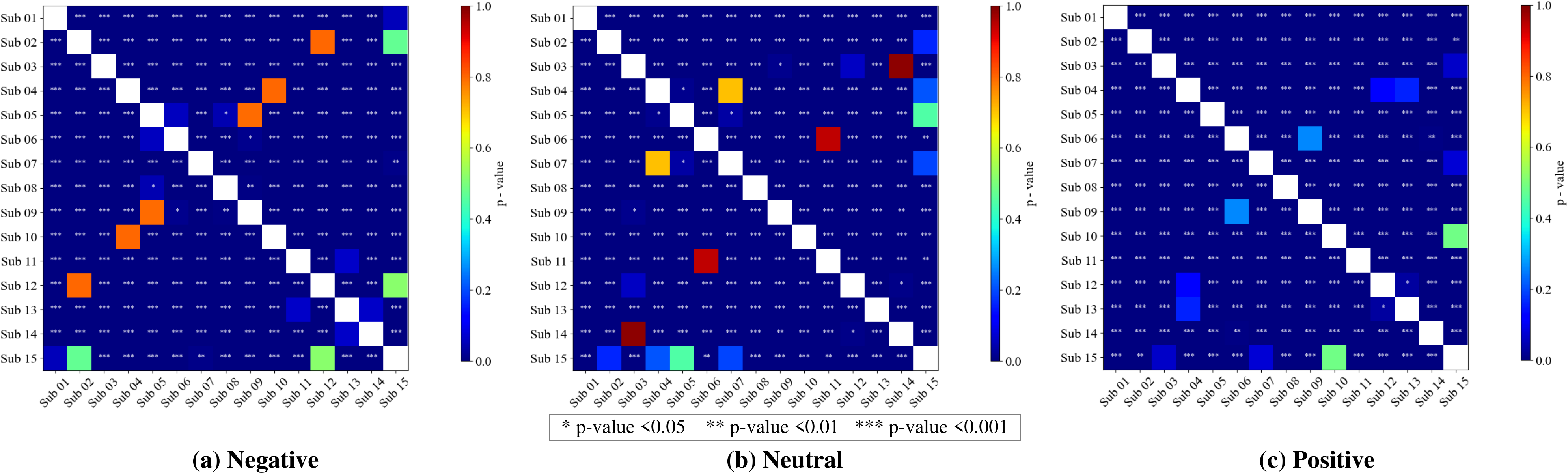}
\caption{Statistical difference matrix across subjects in a single session of SEED database, with independent samples t-test and FDR correction (a) Negative, (b) Neutral, (c) Positive.}
\label{fig:SEED_statistical_matrix}
\end{center}
\end{figure*}

\begin{figure*}[h]
\begin{center}
\includegraphics[width=1\textwidth]{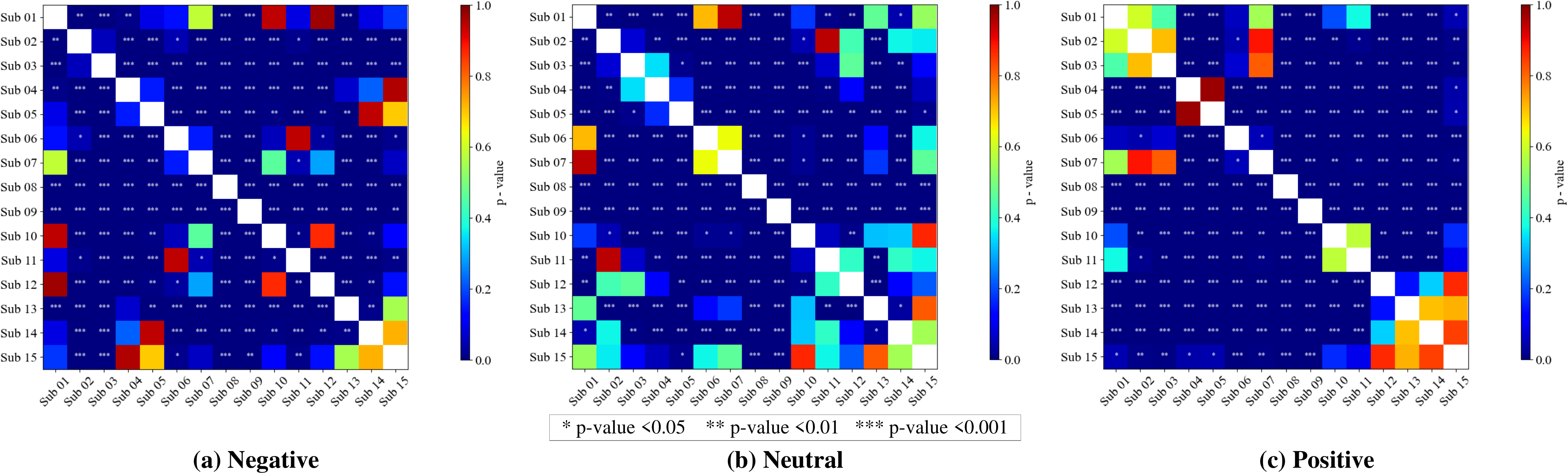}
\caption{Statistical difference matrix across subjects and across session of SEED database, with independent samples t-test and FDR correction (a) Negative, (b) Neutral, (c) Positive.}
\label{fig:SEED_statistical_matrix_3session}
\end{center}
\end{figure*}

\begin{figure}[h]
\begin{center}
\includegraphics[width=1\linewidth]{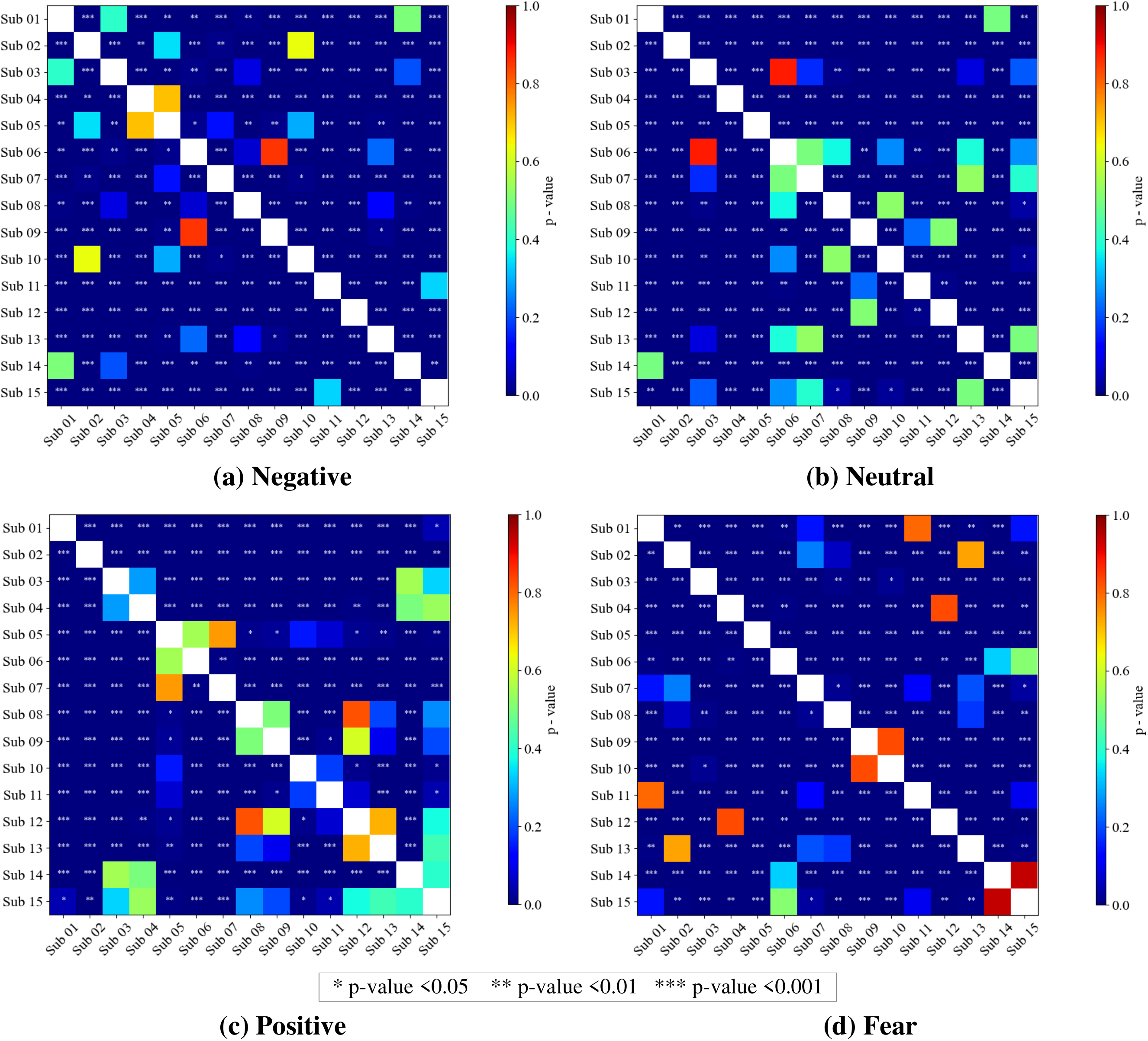}
\caption{Statistical difference matrix across subjects in a single session of SEED-IV database, with independent samples t-test and FDR correction (a) Negative, (b) Neutral, (c) Positive, (d) Fear.}
\label{fig:SEEDIV_statistical_matrix}
\end{center}
\end{figure}

\begin{figure}[h]
\begin{center}
\includegraphics[width=1\linewidth]{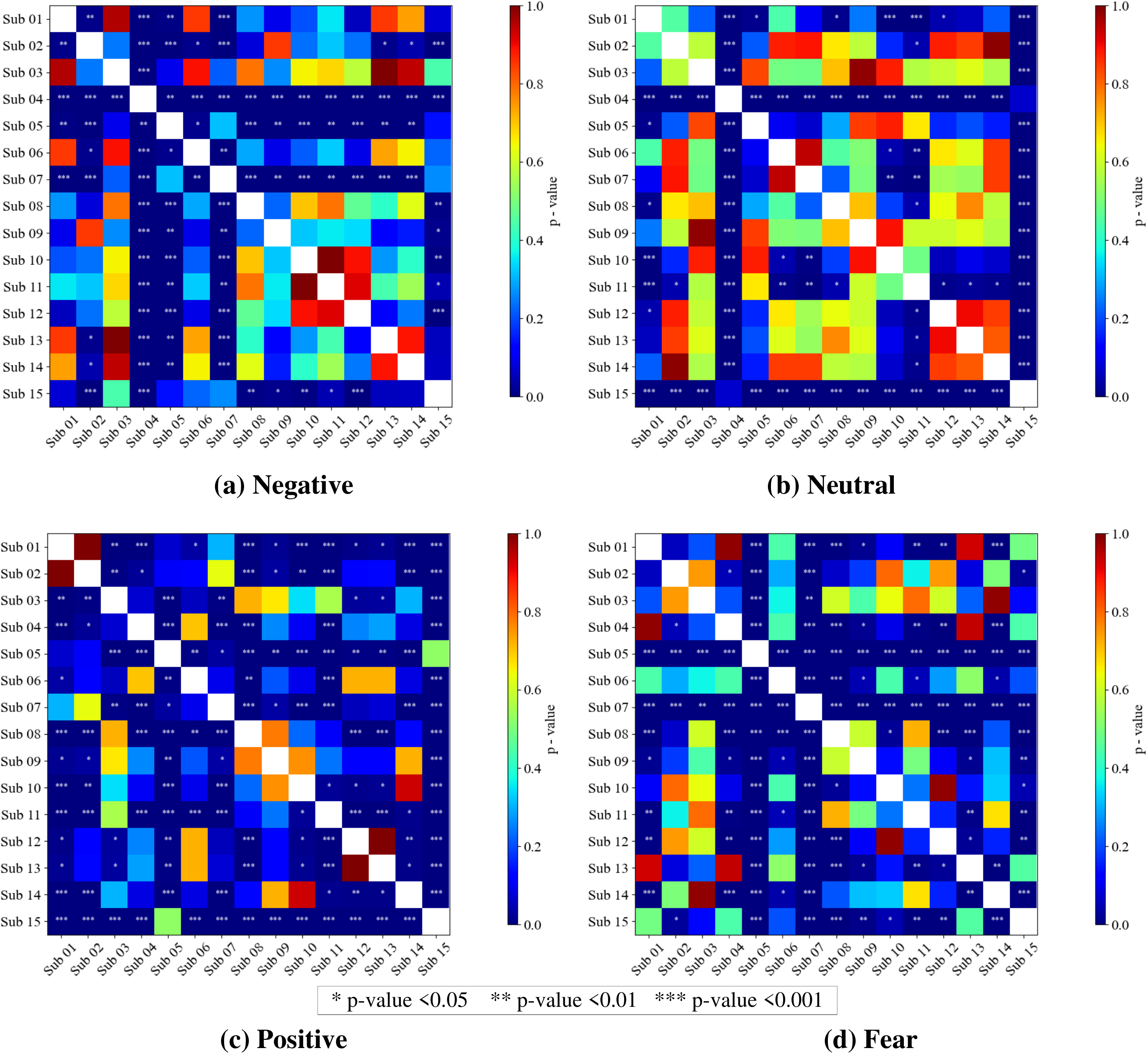}
\caption{Statistical difference matrix across subjects and across session of SEED-IV database, with independent samples t-test and FDR correction (a) Negative, (b) Neutral, (c) Positive, (d) Fear.}
\label{fig:SEEDIV_statistical_matrix_3session}
\end{center}
\end{figure}

Thirdly, to systematically quantify the degree of inter-subject difference, we perform statistical difference matrices using independent sample t-tests (or Wilcoxon rank-sum tests when normality assumptions are violated), with all p-values adjusted using False Discovery Rate (FDR) correction. The resulting matrices (Fig. \ref{fig:SEED_statistical_matrix}-\ref{fig:SEEDIV_statistical_matrix_3session}) indicate statistically significant differences ($p < 0.05$) between subjects under the same emotion condition, underscoring the presence of subject-specific EEG patterns. These inter-individual differences may stem from a range of factors, including variations in physiological structure, psychological characteristics, and individual cognitive processing mechanisms.

In summary, our analysis based on three parts (low-dimensional visualizations using t-SNE, distributional comparisons through box plots, and statistical testing via difference matrices) demonstrates that EEG signals exhibit strong individual variability across subjects and emotional conditions, both visually and statistically. These findings highlight the complexity of cross-subject EEG emotion recognition and emphasize the need for robust adaptation mechanisms such as our proposed M3D framework.

\section{The effect of the parameter}
\begin{figure*}[h]
\begin{center}
\includegraphics[width=1\textwidth]{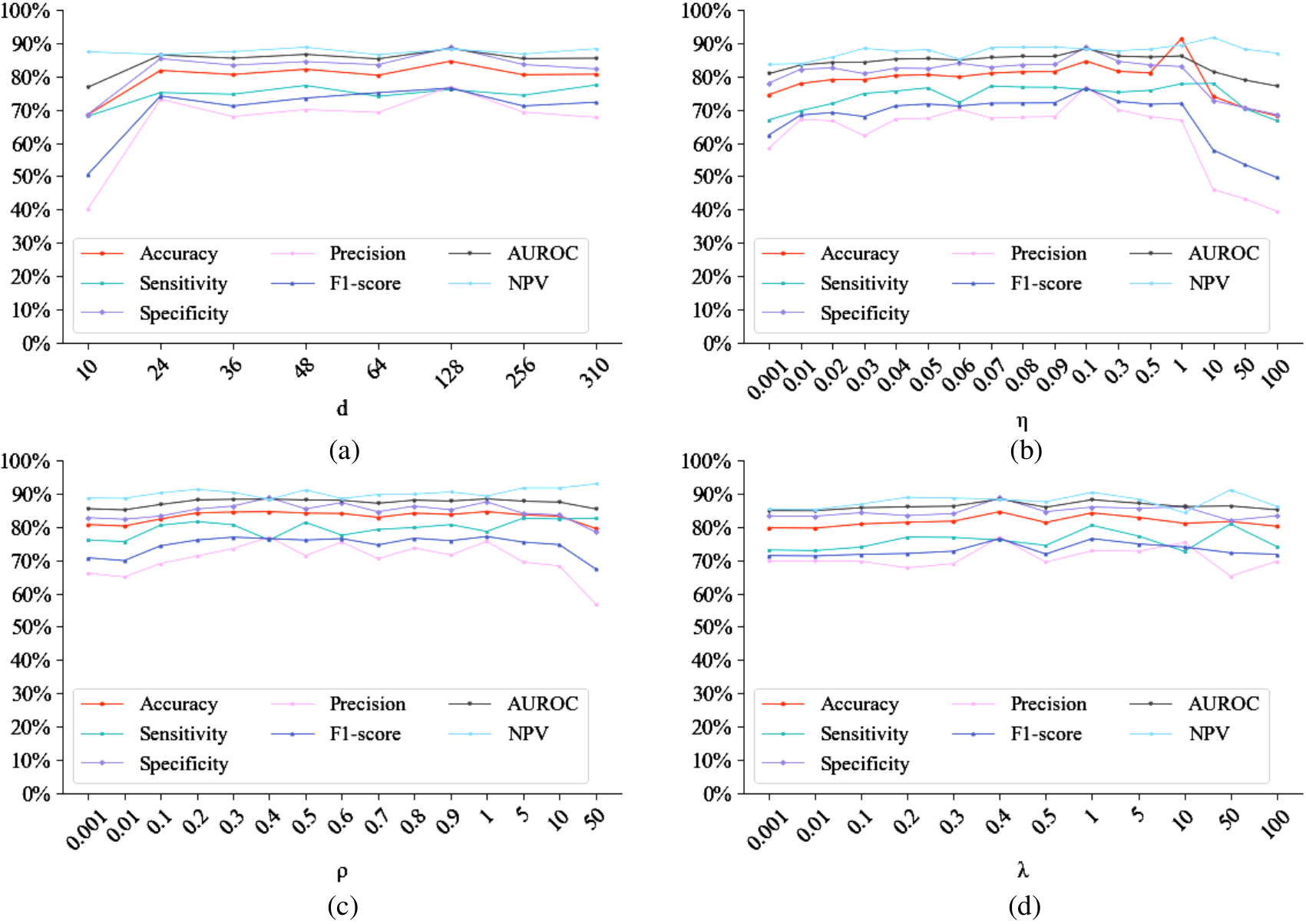}
\caption{Parameter sensitivity and convergence analysis of M3D (a) subspace dimension $d$, (b) regularization parameters $\eta$, (c) regularization parameters $\rho$, (d) regularization parameters $\lambda$.}
\label{fig:parameter}
\end{center}
\end{figure*}

We conduct supplementary experiments under the cross-subject single-session setting using the SEED database to discuss the impacts of key parameters (subspace dimension $d$, regularization parameters $\eta$, $\rho$, $\lambda$), the number of iterations $\iota$, and the training sample size on the performance of the M3D model. The subspace dimension $d$ is varied across $[10,24,36,48,64,128,256,310]$, while the regularization parameters $\eta$, $\rho$, and $\lambda$ are explored at multiple values within the ranges of $[0.001-100]$, $[0.001-50]$, and $[0.001-100]$, respectively. As shown in Fig. \ref{fig:parameter}, the experimental results demonstrate that the M3D model achieves optimal performance across multiple evaluation metrics, when $d$ is set to 128, $\eta$ to 0.1, $\rho$ to 1, and $\lambda$ to 0.4. Additionally, the model maintains stable performance across a broad range of parameter values, particularly when $\eta\in[0.05,1]$, $\rho\in[0.1,1]$, and $\lambda\in[0.1,1]$. These findings confirm the strong parameter adaptability of the proposed M3D model and offer practical guidance for hyperparameter tuning in real-world applications.

Additionally, to examine the classifier learning process, we also investigate how sensitive the model is to changes in the number of iterations ($\iota$) in the learning process. Here, we vary $\iota$ value in a range of $[2,5,10,20,30,50]$. The corresponding learning results, expressed as accuracy rates, are 80.39\%, 83.81\%, 84.57\%, 83.84\%, 83.34\% and 82.55\% respectively. It shows the model's classification performance experiences fluctuations during the initial few iterations but achieves stability beyond $\iota=10$ iterations. This demonstrates the training advantage of M3D in cross-domain tasks.

\begin{figure}[h]
\begin{center}
\includegraphics[width=1\linewidth]{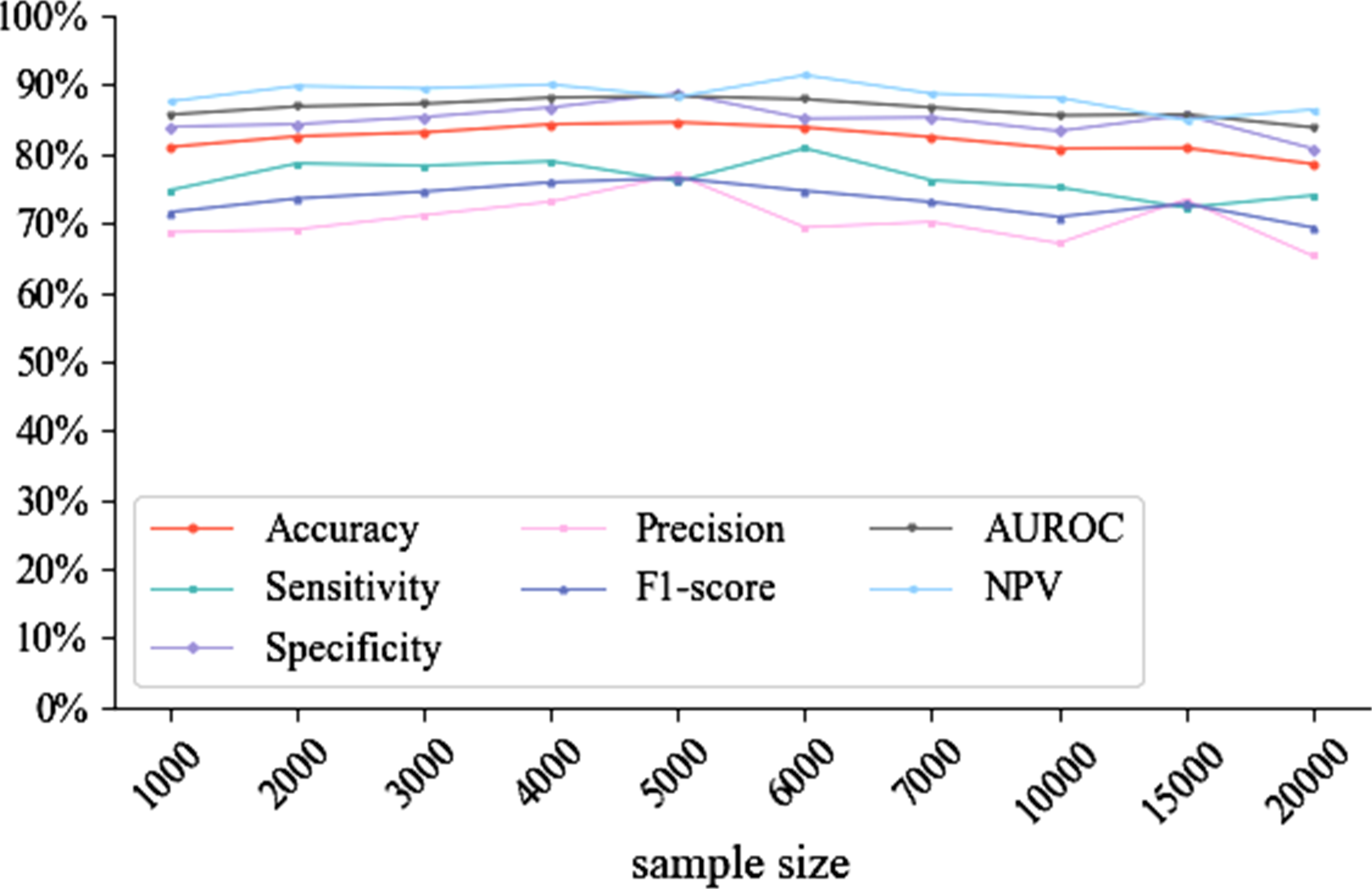}
\caption{The mean accuracies (\%) of the proposed M3D under different sample sizes.}
\label{fig:samplesize}
\end{center}
\end{figure}

We also explore the impact of the training sample size on the model's performance, as well as their computation time. We train the model using training set samples of different scales, with the number of training samples ranging from 1,000 to 20,000. The experimental results in Fig. \ref{fig:samplesize} and Table \ref{tab:samplesize} show that our method achieves an accuracy of 83.11\% with only 3,000 training samples, and the runtime is only 37.4 seconds. Notably, this outperforms several deep learning methods that use all available training samples. For instance, GECNN \cite{pan2023st} achieves an accuracy of 82.46±10.83\%, MS-MDA \cite{she2023multisource} obtains 79.67±8.01\%, and R2G-STLT \cite{cheng2024emotion} reaches 77.96±6.38\%.

\begin{table*}[h]
\setlength{\tabcolsep}{0.45em}
\begin{center}

\caption{Running time (s) of the proposed M3D under different sample sizes.}
\label{tab:samplesize}
\scalebox{1}{
\begin{tabular*}{\hsize}{@{}@{\extracolsep{\fill}}lcccccccccc@{}}
\toprule
Sample size & 1000 &2000&3000&4000 &5000&6000&7000&10000&15000 &20000\\
\midrule
Runnning time (s) &19.32 &27.63 &37.40 &53.35 &81.81&88.51 &100.95 &263.17 &449.41 &1205.00
\\
\bottomrule
\end{tabular*}
}
\end{center}
\end{table*}

\begin{table}

\caption{Performance comparison using different modules.}
\label{tab:seed_singlesession_different_modules}
    \centering
    \normalsize 
    \renewcommand{\arraystretch}{1.2}
    \resizebox{\columnwidth}{!}{%
    \setlength{\tabcolsep}{3.5mm}{
        \begin{tabular}{l|cccc}
        \toprule
        Manifold & \multicolumn{2}{c}{Stiefel manifold}  & \multicolumn{2}{c}{Grassmann manifold}    \\ 
        \midrule
        Accuracy  & \multicolumn{2}{c}{67.94} & \multicolumn{2}{c}{\textbf{84.57}}  \\
        Sensitivity &\multicolumn{2}{c}{63.30}&\multicolumn{2}{c}{\textbf{76.04}} \\
        Specificity &\multicolumn{2}{c}{70.37}&\multicolumn{2}{c}{\textbf{88.78}} \\
        Precision &\multicolumn{2}{c}{52.77}&\multicolumn{2}{c}{\textbf{76.94} }\\ 
        F1-Score & \multicolumn{2}{c}{57.55} &\multicolumn{2}{c}{\textbf{76.49}}\\
        AUROC &\multicolumn{2}{c}{75.95}&\multicolumn{2}{c}{\textbf{88.36}} \\ 
        NPV &\multicolumn{2}{c}{78.57}&\multicolumn{2}{c}{\textbf{88.27}} \\
        \midrule
        \midrule
        Dimensionality Reduction & \multicolumn{2}{c}{PCA}  & \multicolumn{2}{c}{TCA}    \\ 
        \midrule
        Accuracy  & \multicolumn{2}{c}{56.81} & \multicolumn{2}{c}{\textbf{84.57}}  \\
        Sensitivity &\multicolumn{2}{c}{77.73}&\multicolumn{2}{c}{\textbf{76.04}} \\
        Specificity &\multicolumn{2}{c}{50.91}&\multicolumn{2}{c}{\textbf{88.78}} \\
        Precision &\multicolumn{2}{c}{30.86}&\multicolumn{2}{c}{\textbf{76.94} }\\ 
        F1-Score & \multicolumn{2}{c}{44.18} 
        &\multicolumn{2}{c}{\textbf{76.49}}\\
        AUROC &\multicolumn{2}{c}{78.28}&\multicolumn{2}{c}{\textbf{88.36}} \\ 
        NPV &\multicolumn{2}{c}{89.03}&\multicolumn{2}{c}{\textbf{88.27}} \\
        \midrule
        \midrule
        $\mu$  &   0 & 0.5& 1& dynamic   \\ 
        \midrule
        Accuracy  & 78.57 &81.55 &80.31 & \textbf{84.57}  \\
        Sensitivity  &70.89&75.42&71.69&\textbf{76.04}\\
        Specificity &82.80&84.60&84.96&\textbf{88.78}\\
        Precision&69.38 &70.94&72.03&\textbf{76.94}\\ 
        F1-Score &70.13 &73.11 &71.86 &\textbf{76.49}\\
        AUROC &84.12&86.16&85.31&\textbf{88.36}\\ 
        NPV&83.80 &87.35&84.74&\textbf{88.27}\\
        \midrule
        \midrule
        Ensemble Methods & Last & Averaging & Voting & LinkCluE  \\ 
        \midrule
        Accuracy &79.66&78.60&80.62&\textbf{84.57}\\
        Sensitivity &74.37 &79.58&73.38&\textbf{76.04}\\
        Specificity &82.21&78.23&84.38&\textbf{88.78}\\
        Precision&66.86 &57.86&70.94&\textbf{76.94}\\ 
        F1-Score &70.41&67.00&72.14&\textbf{76.49}\\
        AUROC&84.68 &84.17&85.50&\textbf{88.36}\\ 
        NPV &86.92&91.07&85.92&\textbf{88.27}\\
        \bottomrule
        \end{tabular}}}
\end{table}

\section{Theoretical analysis of the computational complexity}
We conduct a comprehensive theoretical analysis of the computational complexity of the proposed M3D framework, taking into account its main modules: manifold feature transformation, dynamic distribution alignment, classifier learning, and ensemble learning.

Let $n$ and $m$ be the sample sizes of the source and target domains respectively, and $d$ be the dimensionality of the reduced features. In the dynamic distribution alignment module, the computational cost is mainly consumed in the dimensionality reduction process (such as Transfer Component Analysis, TCA) and the computation of the geodesic kernel $G$ on the Grassmann manifold. Specifically, the time complexity of the dimensionality reduction is $O(d(n+m)^2)$. Calculating the geodesic kernel $G$ involves matrix operations such as singular value decomposition (SVD), leading to a time complexity of $O(d^3)$. Key computations in the dynamic distribution alignment module include calculating the adaptive factor $\mu$ and constructing the marginal and conditional distribution matrices. The calculation of $\mu$ incurs a complexity of $O(n+m)$, while computing the distribution matrices requires $O(C(n+m)^2)$, where $C$ is the total number of classes. In the classifier learning module, it involves solving an optimization problem through matrix inversion, which leads to a time complexity of $O((n+m)^3 )$. Since this process is iteratively performed $l$ times, the overall complexity of this module is $O(l(n+m+C(n+m)^2+(n+m)^3))$. In the ensemble learning module, the computational cost is dominated by calculating the similarity matrix with $O(m^2)$ complexity, followed by hierarchical clustering with \ensuremath{O(m^2 \log m)} complexity.

By combining the complexities of all modules, the overall time complexity of M3D can be expressed as:

\begin{multline}
    O\bigl(d(n+m)^2 + d^3 + l(n+m + C(n+m)^2 \\
    + (n+m)^3 + m^2 + m^2 \log m\bigr).
\end{multline}
In our experimental setting, considering that $l$, $d$, $C$ and $m$ are relatively small compared to $(n+m)$, the dominant term is $O(l(n+m)^3)$. This shows that the proposed M3D maintains a relatively low computational complexity compared to deep learning models, which typically involve repeated backpropagation and large-scale parameter optimization.

\section{Investigating factors affecting model performance}
We also conduct a series of experiments to study the factors that affect the performance of the model. To evaluate the efficacy of the Grassmann manifold, we conduct a comparative experiment by replacing it with the Stiefel manifold in the manifold feature transformation module. The Grassmann manifold excels at depicting the geometric structure of subspaces in high-dimensional data, precisely capturing the internal connection between EEG signals and emotions, while the Stiefel manifold mainly focuses on the manifold composed of orthogonal matrices and lacks the specific ability to extract EEG emotion features. As shown in Table \ref {tab:seed_singlesession_different_modules}, after the replacement, the accuracy drops to 67.94\%, and the average performance of 7 indicators decreases by 16.14\%. This further demonstrates that the Grassmann manifold has significant advantages in extracting EEG features related to emotions, which is more conducive to improving the model's emotion recognition performance.

To evaluate the efficacy of TCA in the manifold feature transformation, we assess the model performance by comparing the use of TCA against substituting TCA with Principal Component Analysis (PCA). As shown in Table \ref{tab:seed_singlesession_different_modules}, the model exhibits significantly better performance in terms of accuracy and F1-Score when comparing TCA with PCA. This improvement can be attributed to TCA's ability to enhance model performance by simultaneously mapping and reducing dimensions, and by maximizing the similarity between datasets from two domains. It would facilitate more effective dynamic distribution alignment and classifier categorization.

To evaluate the effectiveness of dynamic distribution alignment, we opt for a fixed $\mu$ value approach, rather than using an adaptive $\mu$. The $\mu$ values are set at $0$, $0.5$, and $1$. At $\mu=0$, the model exclusively focuses on the marginal distribution. With $\mu=0.5$, both marginal and conditional distributions are considered equally. At $\mu=1$, the model solely emphasizes the conditional distribution. As reported in Table \ref{tab:seed_singlesession_different_modules}, it shows that employing an adaptive $\mu$ can significantly enhance model performance, resulting in a 4.26\% improvement in accuracy. The results demonstrate that dynamic distribution alignment can effectively quantify the disparity between source and target domains and accurately evaluate the significance of the differences between marginal and conditional distributions. This capability is particularly advantageous for the classification performance, which is tailored for addressing variations between distributions.

To evaluate the impact of ensemble learning on model performance, we adjust our approach by incorporating various ensemble learning techniques, namely the averaging method, the voting method, and the LinkCluE method. Note here that LinkCluE is the method implemented in the proposed M3D. Based on the obtained classification results in the classifier learning ($\hat{y}_t^{(\iota)}$ ($\iota=1,\cdots,10$)), the effectiveness of ensemble learning is analyzed. Furthermore, we draw comparisons to scenarios devoid of ensemble learning strategies by directly employing the final classification outcome from the last iteration loop in the classifier learning, denoted as $\hat{y}_t^{(\iota)}$ (with $\iota=10$), which serves as our baseline. The experimental comparison results are reported in Table \ref{tab:seed_singlesession_different_modules}. It shows that ensemble learning with voting and LinkCluE could enhance the reliability of classification results and improve the overall performance. This confirms the effectiveness of ensemble algorithms. Furthermore, the LinkCluE method outperforms the voting method, demonstrating a superior ability to synthesize classification outcomes across iterations and to discern the underlying patterns of the classification results.

\section{Visualization of Feature Alignment}
\begin{figure}[h]
\begin{center}
\includegraphics[width=1\linewidth]{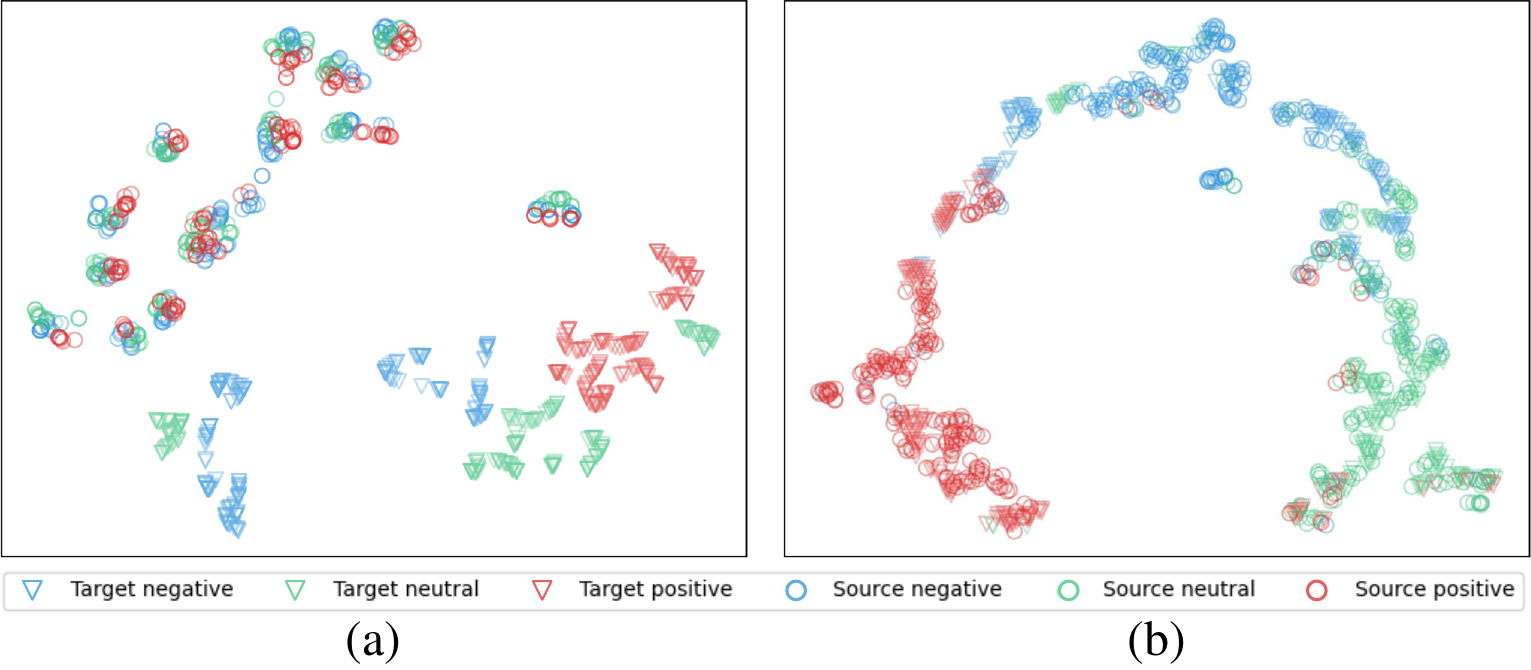}
\caption{A visualization of the learned feature representations (a) before alignment, (b) after alignment. Here, the circle and triangle represent the source domain ($S$) and the target domain ($T$). The red, blue and green colors indicate happy, sad and neural emotions.}
\label{fig:SEED_TSNE}
\end{center}
\end{figure}

We employ t-SNE visualizations and EEG topographic maps to intuitively demonstrate the alignment of source and target features before and after adaptation. Specifically, we utilize the t-SNE algorithm \cite {van2008visualizing} to visualize the source and target features from the SEED database both before and after alignment. The results are illustrated in Fig. \ref{fig:SEED_TSNE}. To enhance clarity, we assign different colors to the samples with different emotions and assign different shapes to distinguish samples from different domains. The results demonstrate that M3D reduces the distribution discrepancy among the two domains throughout the training process. And it progressively separates the samples with different emotions to minimize the emotion classification error. In summary, the results provide supplementary evidence to support the efficacy of the proposed M3D in EEG-based emotion.

In the topographic analysis, we identify the important brain patterns for emotion recognition by computing the mutual information between the brain patterns and prediction labels. The EEG topographic maps can intuitively display the spatial distribution characteristics of neural activities related to different emotions. Specifically, we identify the important brain patterns for emotion recognition by computing the mutual information between the brain patterns and prediction labels. We have the data input of the target domain, given as $X_{t}$, which is a $M\times310$ matrix. Here, $M$ is the sample size in the target domain, and 310 refers to the extracted DE features from 62 channels at 5 frequency bands. The corresponding model prediction result is $\hat{Y_{t}}$, with a size of $M\times3$. The columns in $\hat{Y_{t}}$ indicate the prediction probabilities of different emotions (negative, neutral, and positive). Then, we estimate the mutual information between the input features $X_{t}$ and and the prediction result $\hat{Y_{t}}$ using the non-parametric method as stated in \cite{1987sample,kraskov2004estimating,ross2014mutual}. The obtained mutual information matrix is termed as $I(X_{t},\hat{Y_{t}})\in{R^{3\times310}}$, indicating a quantification of the inherent dependence between the EEG patterns and the model prediction results. $I(X_{t},\hat{Y_{t}})$ is further normalized to $[0,1]$, and a larger value refers to a greater informativeness of the EEG patterns to the model prediction. For the $\hat{Y_{t}}$ before alignment, it is the prediction result obtained by directly inputting the DE features into the DT classifier. While for the $\hat{Y_{t}}$ after alignment, it is the final result obtained by inputting the DE features into the model M3D. 

\section*{References}
\bibliographystyle{IEEEtran}
\bibliography{references}
\end{document}